\documentclass[fleqn,usenatbib]{mnras}

\usepackage{newtxtext,newtxmath}
\usepackage{threeparttable}

\usepackage[T1]{fontenc}

\DeclareRobustCommand{\VAN}[3]{#2}
\let\VANthebibliography\thebibliography
\def\thebibliography{\DeclareRobustCommand{\VAN}[3]{##3}\VANthebibliography}

\usepackage{graphicx}	
\usepackage{amsmath}	

\title[The Origin of Photospheric Emission of GRB 220426A]{The Origin of the Photospheric Emission of GRB 220426A}

\author[Xin-Ying Song et al.]{
Xin-Ying Song,$^{1}$\thanks{E-mail: songxy@ihep.ac.cn}
Shuang-Nan Zhang,$^{1,2}$
Ming-Yu Ge,$^{1}$
Shu Zhang$^{1}$
\\
$^{1}$Key Laboratory of Particle Astrophysics, Institute of High Energy Physics, Chinese Academy of Sciences, Beijing 100049, China\\
$^{2}$University of Chinese Academy of Sciences, Chinese Academy of Sciences, Beijing 100049, China\\
}

\date{Accepted 2022 September 20. Received 2022 August 16; in original form 2022 August 16}

\pubyear{2022}

\begin{document}
\label{firstpage}
\pagerange{\pageref{firstpage}--\pageref{lastpage}}
\maketitle

\begin{abstract}
GRB 220426A is a bright gamma-ray burst (GRB) dominated by the photospheric emission. We perform several tests to speculate the origin of this photospheric emission. The dimensionless entropy $\eta$ is large, which is not usual if we assume that it is a pure hot fireball launched by neutrino-antineutrino annihilation mechanism only. Moreover, the outflow has larger $\eta$ with lower luminosity $L$ in the first few seconds, so that the trend of time-resolved $\eta-L$ can not be described as a monotonically positive correlation between $\eta$ and $L$. A hybrid outflow with almost completely thermalized Poynting flux could account for the quasi-thermal spectrum as well as large $\eta$. More importantly, the existence of magnetic field could affect the proton density and neutron-proton coupling effect, so that it could account for the observed trend of time-resolved $\eta-L$. The other origins for the photospheric emission, such as non-dissipative hybrid outflow or magnetic reconnection, are not supported because their radiation efficiencies are low, which is not consistent with non-detection of the afterglow for GRB 220426A. Therefore, we think the hybrid outflow may be the most likely origin.
\end{abstract}




\section{Introduction} \label{sec:intro}

The fireball model was first suggested by \cite{ 1978MNRAS.183..359C}, \cite{1986Are} and \cite{1986ApJ308L43P}. As a prediction of the fireball model, the photospheric emissions in different regimes have been discussed in some works \citep[e.g.,][]{1994ApJ...430L..93R,2000ApJ...530..292M}. Although quasi-thermal GRBs are not usual, there are some cases of study, such as GRB 090902B, GRB 210121A, and recent GRB 220426A~\cite[e.g.,][]{2009ApJ...706L.138A,2021ApJ...922..237W,2022arXiv220409430S,2022arXiv220508427W,2022arXiv220508737D}. \cite{2008ApJ...682..463P} shows that photons make their last scatterings at a distribution of radii and angles. A probability photosphere model is introduced~\citep[e.g.,][]{2008ApJ...682..463P,2013A,Deng_2014} in the case of a pure hot fireball, in which the Planck spectrum is modified by geometrical broadening and could be described by a multi-blackbody (mBB) spectrum~\citep[e.g.,][]{2010ApJ...709L.172R,Hou_2018}. 
  Based on this model, the theory of non-dissipative photospheric (NDP) emission from relativistic jets with angle-dependent outflow properties is developed in ~\cite{2013A}, and a structured jet with angle-dependent baryon loading parameter profiles is considered. Average low-energy photon index ($\alpha=-1$) could be obtained in saturated regime and independent of viewing angle.%

Dissipations below the photosphere can heat electrons above the equilibrium temperature. These electrons emit synchrotron emission and comptonize thermal photons, thereby also modify the shape of the Planck spectrum~\citep{2005ApJ...635..476P, Pe_er_2006, 2005Dissipative}. The observational evidence for the subphotospheric heating has been provided by \cite{2011Observational}. Besides, internal shocks below the photosphere~\citep{2005Dissipative}, magnetic reconnection~\citep{1994A, 2004Spectra}, and hadronic collision shocks~\citep{2010Collisional, 2010Radiative} can also cause dissipation. Especially, magnetic dissipation may occur during the propagation of the jet below the photosphere and cause 
a non-thermal appearance~\citep[e.g.,][]{2006A&A...457..763G, 2008A&A...480..305G,2015ApJ...801..103G}. Particularly, if the magnetization complete thermalization below the photosphere is likely to be achieved and magnetic energy is directly converted to the heat below the photosphere, the spectrum is still the same as that for the non-dissipative case of pure thermal outflow~\citep{2022MNRAS.509.6047M}.

\begin{figure*}
\begin{center}
 \centering
\includegraphics[width=0.75\textwidth]{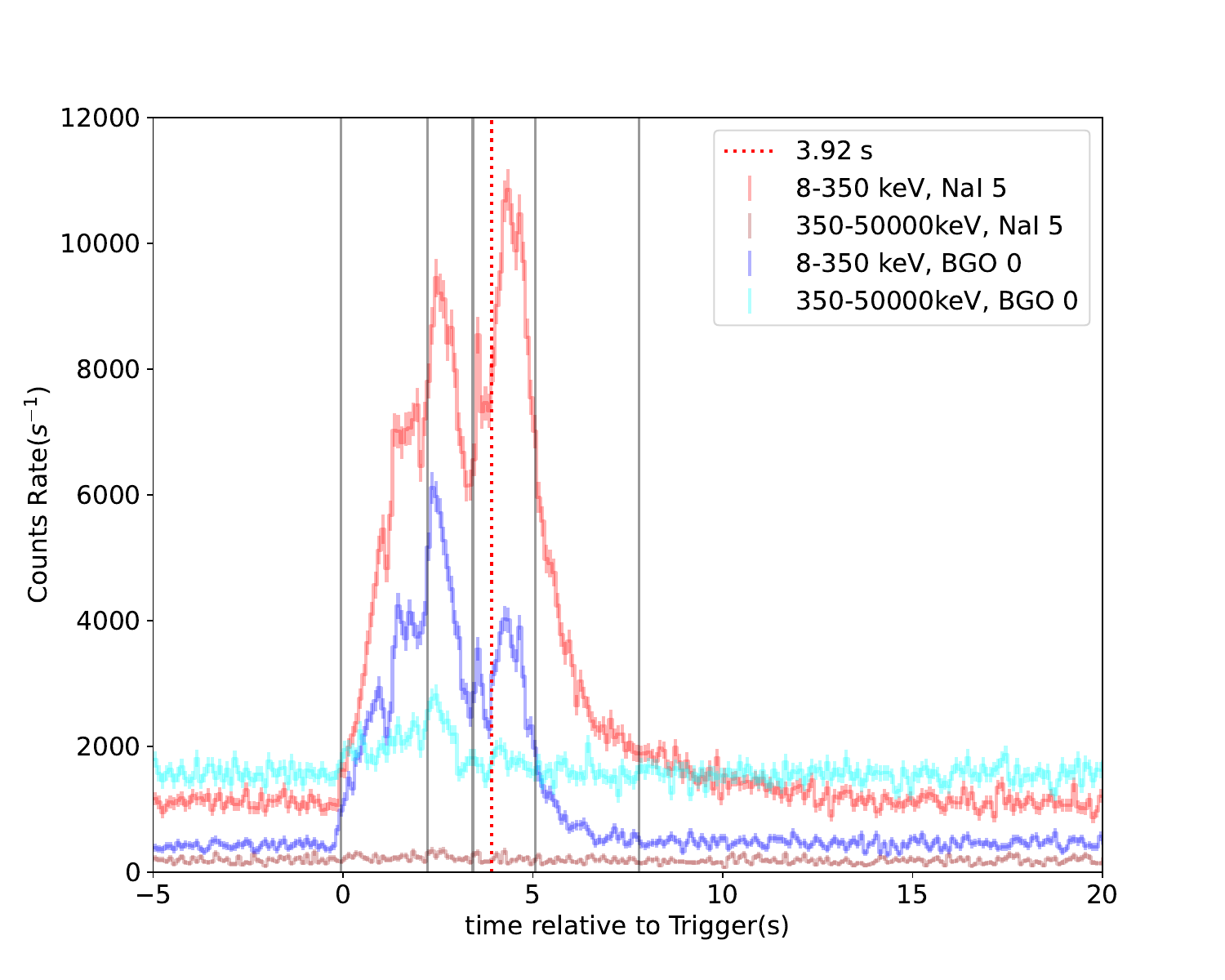}
\caption{ The light curves and time-resolved regions of GRB 220426A. The red, blue lines denote the light curves of NaI 5 and BGO 0 detectors in 8-350 keV, while the brown and cyan lines denote the light curves of NaI 5 and BGO 0 detectors in 350-50000 keV. The black vertical lines represent the time-resolved regions. 
\label{fig:LC} }
\end{center}
\end{figure*}

GRB 220426A is first observed at 06:49:51.230 UT ($T_0$) on 26 April 2022 by Fermi/GBM~\citep[GCN,][]{2022GCN.31955....1M}. The light curves of one brightest NaI and one brightest BGO detector are shown in Figure~\ref{fig:LC}. In this analysis, we perform spectral fits with different models, including the empirical models, such as BAND function, exponential cut-off power law (CPL, also called Comptonized
model), and physical models, such as mBB model, the NDP model in the case of pure hot fireball (PHF)~\citep[e.g.,][]{2008ApJ...682..463P,2013A,Deng_2014} with considering different regimes and the jet structure~\cite[e.g., studies on GRB 170817A,][]{2017MNRAS.471.1652L,2018MNRAS.475.2971B}. We find that the photospheric emission is dominant in prompt emission phase and discuss the origin of this photospheric emission.

This paper is organized as follows: in Section~\ref{sec:modeling}, empirical models as well as physical models are used in time-integrated spectral fit. The general emission properties are extracted and discussed;
in Section~\ref{sec:tests}, several tests are performed to speculate the origin of the emission. The origin of the photospheric emission is discussed; a summary and conclusion are given in Section~\ref{sec:conclusion}.

\section{Basic Spectral modeling and properties of GRB 220426A}\label{sec:modeling}
The data are from one brightest NaI detector (NaI 5) and one brightest BGO detector (BGO 0) of Fermi/GBM. To obtain the spectrum of prompt emission, a polynomial is applied to fit all the energy channels and then interpolated into the signal interval to yield the background photon count estimate for the GRB data. Five models, including CPL, BAND, mBB, NDP in PHF, and synchrotron model are used in modeling the time-integrated spectrum in $[-0.05, 7.79]$ s relative to $T_0$. The details of these models are described in Appendix~\ref{sec:4funcs}. The Markov Chain Monte Carlo (MCMC) fitting is performed to find the parameters with the maximum Poisson likelihood. 
\begin{figure*}
\begin{center}
 \centering
   \includegraphics[width=0.50\textwidth]{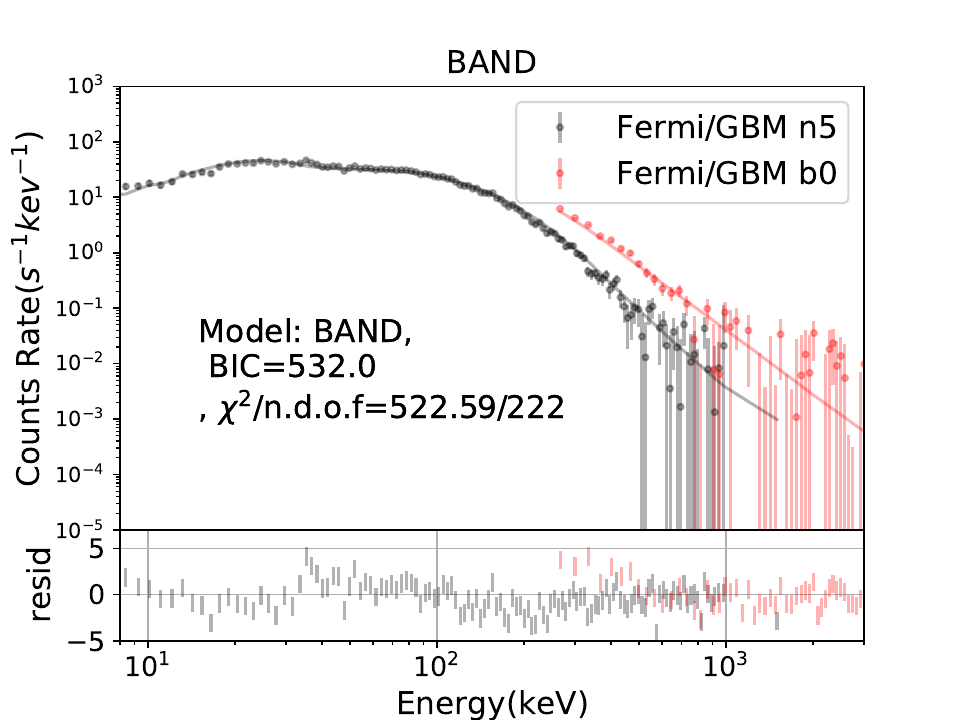}
   \put(-220,160){(a)}
   \includegraphics[width=0.50\textwidth]{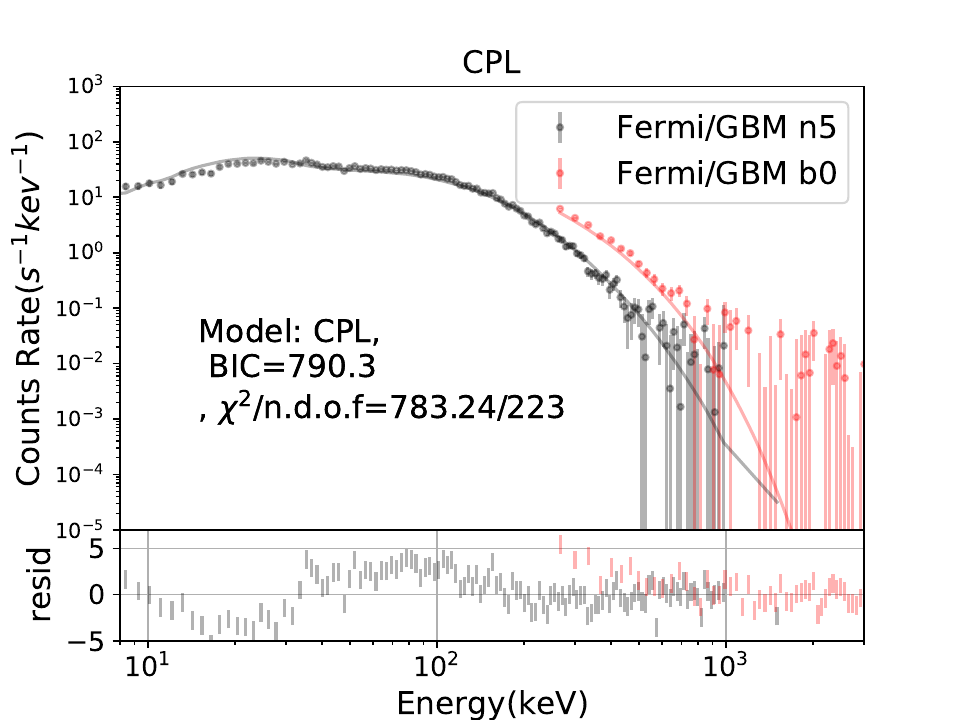}
   \put(-220,160){(b)}\\
   \includegraphics[width=0.50\textwidth]{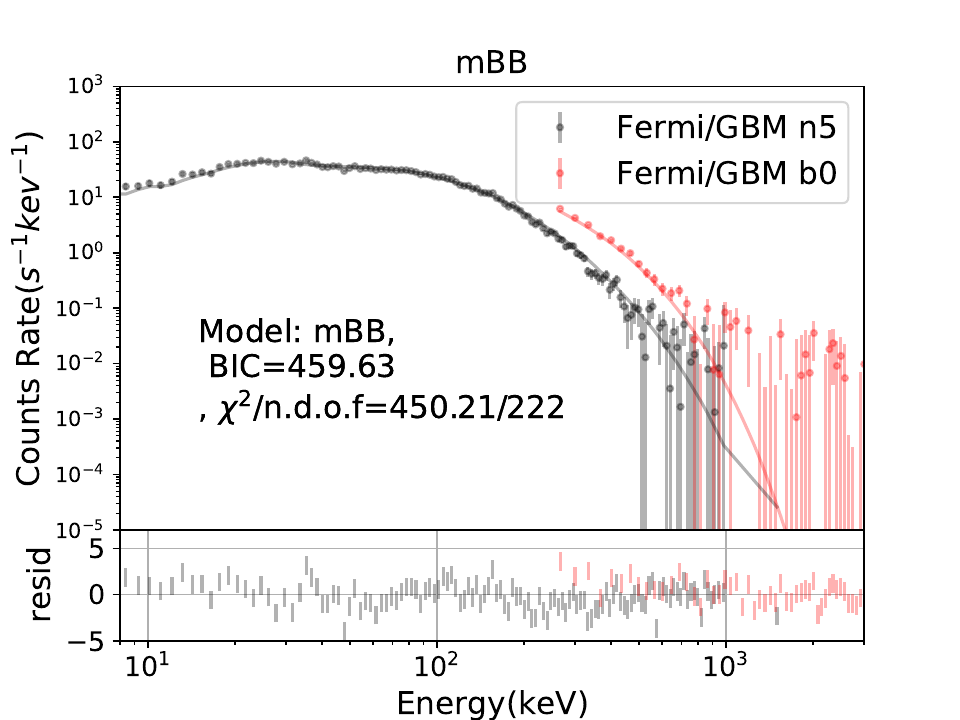}
   \put(-220,160){(c)}
   \includegraphics[width=0.50\textwidth]{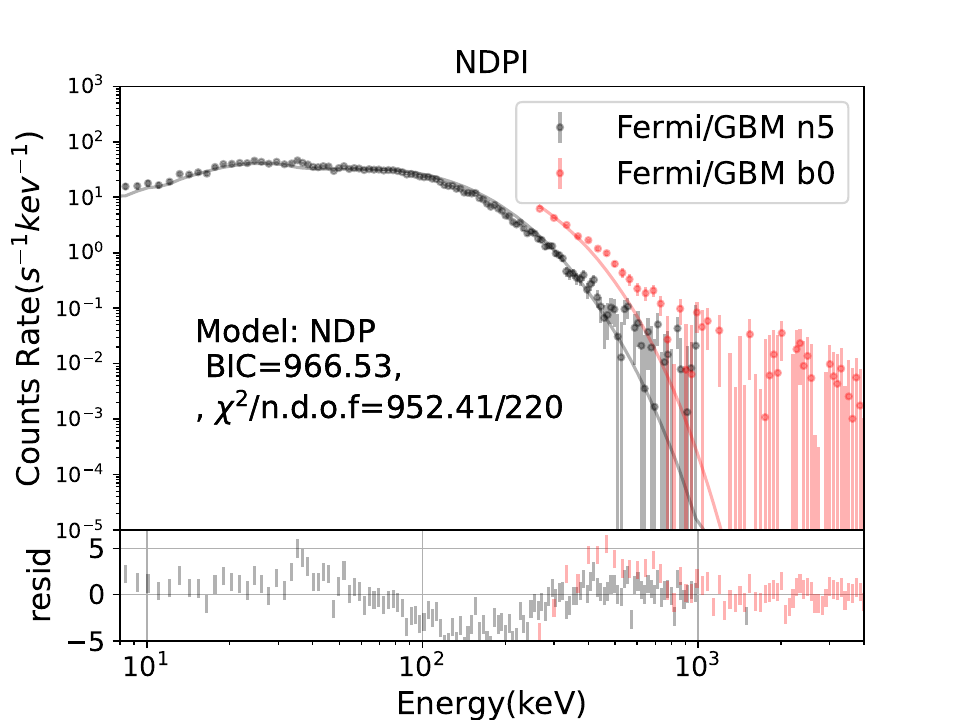}
   \put(-220,160){(d)}\\
   \includegraphics[width=0.50\textwidth]{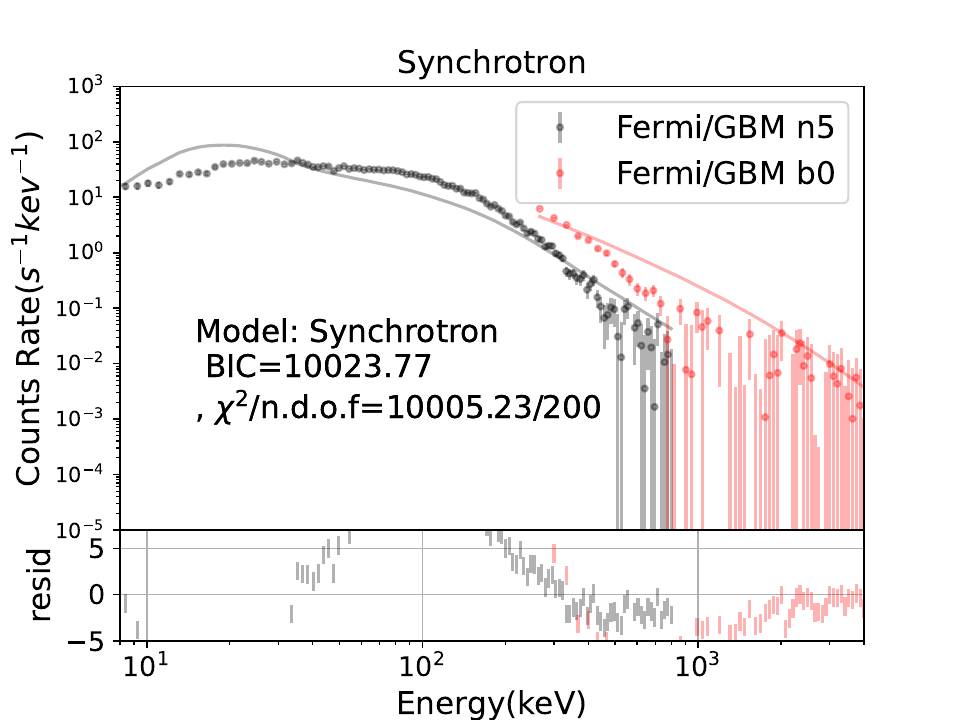}
   \put(-220,160){(e)}
   \includegraphics[width=0.50\textwidth]{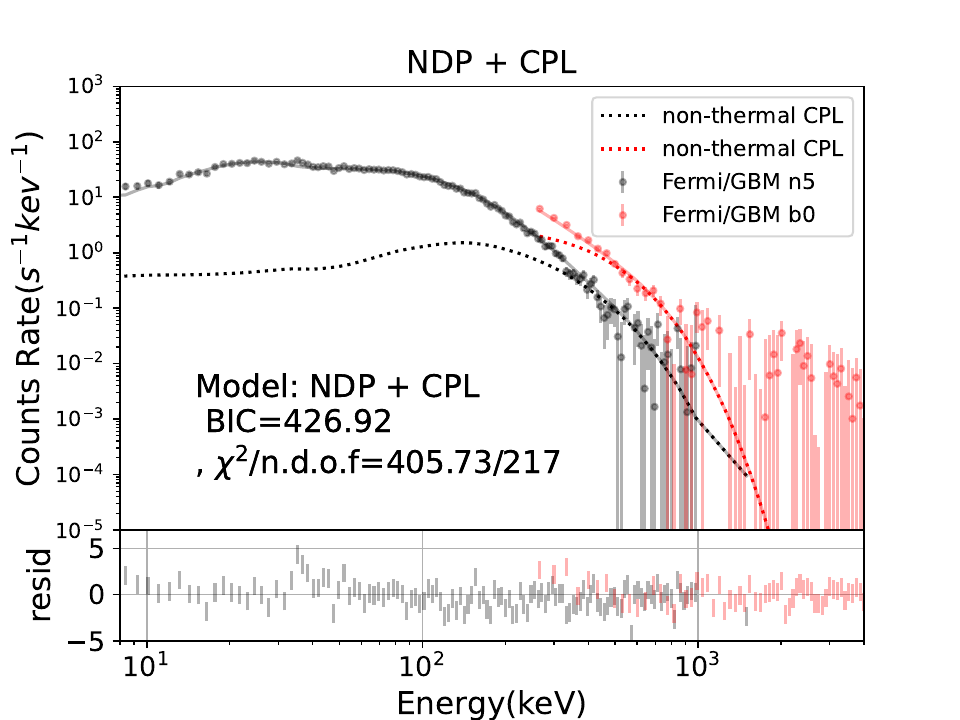}
   \put(-220,160){(f)}
\caption{(a)-(f): Time-integrated spectral fit results in [-0.05,7.79] s with BAND, CPL, mBB, NDP in PHF, synchrotron and NDP +CPL model. 
\label{fig:basic_spectralfit} }
\end{center}
\end{figure*}
Figure~\ref{fig:basic_spectralfit} shows the spectral fit results. With CPL model, $\alpha=-0.14^{+0.01}_{-0.01}$, $E_{\rm p}=162.3^{+1.2}_{-1.3}$ keV and flux is $1.13^{+0.06}_{-0.06}\times10^{-5}$ erg cm$^{-2}$ s$^{-1}$   (bolometric, the same below). With Band model, $\alpha=-0.17^{+0.03}_{-0.03}$, $\beta=-3.76^{+0.13}_{-0.16}$, $E_{\rm p}=168.2^{+2.0}_{-2.1}$ keV and flux is $1.36^{+0.13}_{-0.13}\times10^{-5}$ erg cm$^{-2}$ s$^{-1}$. With mBB model, $m=-0.39^{+0.07}_{-0.08}$, $kT_{\rm min}=11.5^{+0.4}_{-0.5}$ keV, $kT_{\rm max}=95.1^{+3.0}_{-2.6}$ keV and flux is $1.30^{+0.09}_{-0.08}\times10^{-5}$ erg cm$^{-2}$ s$^{-1}$. $\alpha$ extracted from BAND or CPL is much greater than -2/3~\citep[so-called `line of death',][]{Preece_1998} of GRB spectra, which implies a photospheric origin. \cite{2016MNRAS.463.1144W} suggested a Bayesian information criterion (BIC) as a tool for model selection; a model that has a lower BIC value than the other is preferred, and if $\Delta$BIC is above 10, the preference is very strong. With this method, mBB is determined to be the best model among BAND, CPL, NDP in PHF and synchrotron models, which implies that thermal component is dominant. Moreover, from Figure~\ref{fig:basic_spectralfit} (e), the spectral shape is significantly different from that of synchrotron model, and the synchrotron origin for prompt emission could be excluded.

Furthermore, there is no positive detection of the afterglow~\cite[GCN,][]{2022GCN.31966....1D} with Swift/XRT and UVOT follow-up observations, \textit{except for an uncataloged X-ray source in the field}.
According to \cite{2005SSRv..120..165B}, the sensitivity of Swift/XRT is $2\times10^{-14}$ erg cm$^{-2}$ s$^{-1}$ in $10^4$ s. Some works \cite[e.g.,][]{2022arXiv220508427W}, estimated the redshift $z\sim1.40$ for GRB 220426A with Amati-relation~\citep[]{2002A&A...390...81A}. Therefore, the luminosity upper limit of the afterglow is $\sim3\times10^{44}$ erg s$^{-1}$. This implies that the kinetic energy of the outflow is rather small, providing further support that the thermal component is dominant in outflow of GRB 220426A. The radiative efficiency ($\epsilon_{\gamma}$) is defined as $\frac{E_{\gamma}}{E_{\gamma} +E_{\rm k}}$, where $E_{\gamma}$ is the radiated energy, and $E_{\rm k}$ is the kinematic energy which could be detected in the afterglow phase. We could speculate that $\epsilon_{\gamma}$ of GRB 220426A should be quite large. 

From the spectra in Figure~\ref{fig:basic_spectralfit} (d) and (f) with NDP and NDP+CPL models, it seems that there is inconsistency in higher energy band ($\gtrsim$300 keV) between NDP model and the observed spectrum. With a combination model of NDP + CPL, as shown in Figure~\ref{fig:basic_spectralfit} (f), the fitting is greatly improved. $\Delta$BIC between NDP+ CPL and mBB is 44.5 with the changes of degree of freedom (dof) equal to 5, thus it seems that the model of NDP+CPL is better than mBB. However, this is evidence only from statistics. From the fit result with mBB shown in Figure~\ref{fig:basic_spectralfit} (c), the so-called inconsistency is not so evident as that only with NDP. Due to the statistic uncertainty of high energy band, it is difficult to clarify whether the inconsistency is from a peaking bump from another emission mechanism or just the difference on the shape between the NDP model and the observed spectrum in high energy band. More physical tests are necessary to clarify the nature.

\section{Tests for the photospheric emission}\label{sec:tests}
 In this section, there are four possible hypotheses discussed in this section: 
\begin{itemize}
\item [\textbf{A.}] The outflow is a pure hot fireball; 
\item[\textbf{B.}]The outflow is hybrid with Poynting flux which is almost completely thermalized. Magnetic energy is converted to heat below the photosphere, therefore, it is a similar case to \textbf{A.} and produce quasi-thermal emission; 
\item[\textbf{C.}] The outflow is hybrid. The magnetic energy is almost only converted into the kinetic energy of the bulk motion, rather than dissipated; 
\item[\textbf{D.}] The outflow is hybrid. Poynting flux in the hybrid outflow is not completely thermalized. There may exists magnetic dissipation, such as internal-collision-induced magnetic reconnection or turbulence (ICMART). 

\end{itemize}
\subsection{A. hypothesis of a pure hot fireball}\label{sec:a.}
In this section, we use reduction to absurdity to test the reasonability of this hypothesis. If the parameters obtained from the fit are consistent with the properties of PHF, it should be considered as an acceptable hypothesis, otherwise, it is not supported.
Besides of extracted fireball properties, the correlation of dimensionless
entropy $\eta$ and the outflow luminosity $L$ is a test of interest. If we take the hyper-accreting black hole (BH) as the central engine, the GRB jet may be launched through two mechanisms. One is $\nu\overline{\nu}$ annihilation in a neutrino-dominated accretion flow (NDAF) \citep[e.g.][]{1999ApJ...518..356P,2001ApJ...557..949N,Di_Matteo_2002,2002ApJ...577..311K,2006ApJ...643L..87G,2007ApJ...657..383C,2007ApJ...664.1011J,2009ApJ...700.1970L,2010A&A...516A..16L,L_2012}, and generates a fireball which is dominated by the thermal component. 
As concluded in \cite{2008MNRAS.390..781M}, \cite{2011MNRAS.410.2302Z}, \cite{L_2012} and \cite{2013ApJ...765..125L}, neutrino annihilation power is described as $L_{\nu\overline{\nu}}\propto \dot{m}^{9/4}$ and the mass loss rate as $\dot{M}_{\nu\overline{\nu}}\propto \dot{E}^{5/3}_{\nu}$, where $\dot{m}$ is the accretion rate; $\dot{E}_{\nu}$ is the total neutrino power. In the neutrino dominated accretion flow, $\dot{E}_{\nu}\propto \dot{m}$, $\dot{M}_{\nu\overline{\nu}}\propto \dot{m}^{5/3}$ and $L_{\nu\overline{\nu}}\propto \dot{m}^{9/4}$, thus, $\eta=L_{\nu\overline{\nu}}/\dot{M}_{\nu\overline{\nu}}c^2\propto L^{7/27}_{\nu\overline{\nu}}$. In the case of a pure hot fireball, the measured total luminosity of the fireball $L\simeq L_{\nu\overline{\nu}}$. For the measurement of time-resolved $\eta-L$ during a burst, other parameters should be taken as constant within the burst, and the index should be 7/27. Note that in the unsaturated regime, the bulk Lorentz factor of outflow can not be accelerated to the value of $\eta$; in the following, we denote the final accelerated bulk Lorentz factor as $\Gamma$, to be distinguished from $\eta$.

To obtain the $\eta-L$ relation, a time-resolved analysis is performed. Two binning methods are applied. (1) Bayesian blocks (BBlocks) method~\citep{2013ApJ...764..167S} with a false alarm probability
$p_0 = 0.01$; we also require the signal-to-noise ratio (S$/$N)~$\geq$~35 at least in one detector, so we combine some adjacent bins. The fine time-resolved analysis are performed in [-0.05, 2.22] s, [2.22, 3.42] s, [3.42, 3.92] s, [3.92, 5.07] s and [5.07, 7.79] s. These five bins are divided by the vertical lines in Figure~\ref{fig:LC}. (2) Constant cadence (CC) method with bin width of 1.0 s; \cite{2014On} suggested that the constant cadence (CC) method is accurate when the cadence is not too coarse. We find that the results are not sensitive to the particular binning method applied.

\begin{figure*}
\begin{center}
 \centering
    \includegraphics[width=0.5\textwidth]{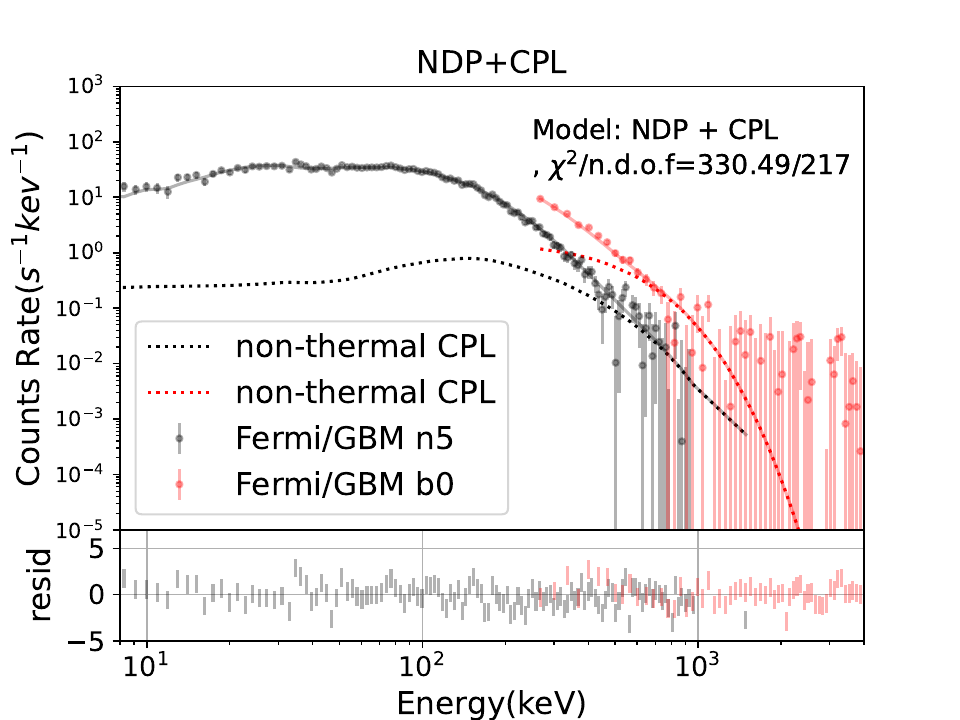}\put(-220,160){(a)[-0.05,3.92]s: NDP+CPL}
    \includegraphics[width=0.5\textwidth]{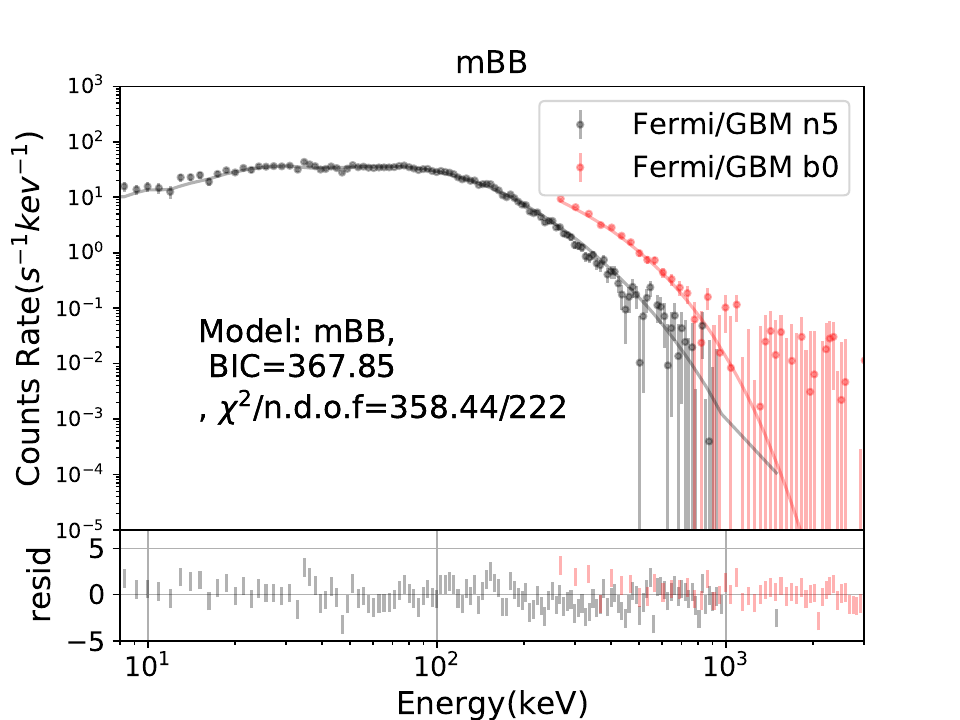}\put(-220,160){(b)[-0.05,3.92]s: mBB }\\   \includegraphics[width=0.75\textwidth]{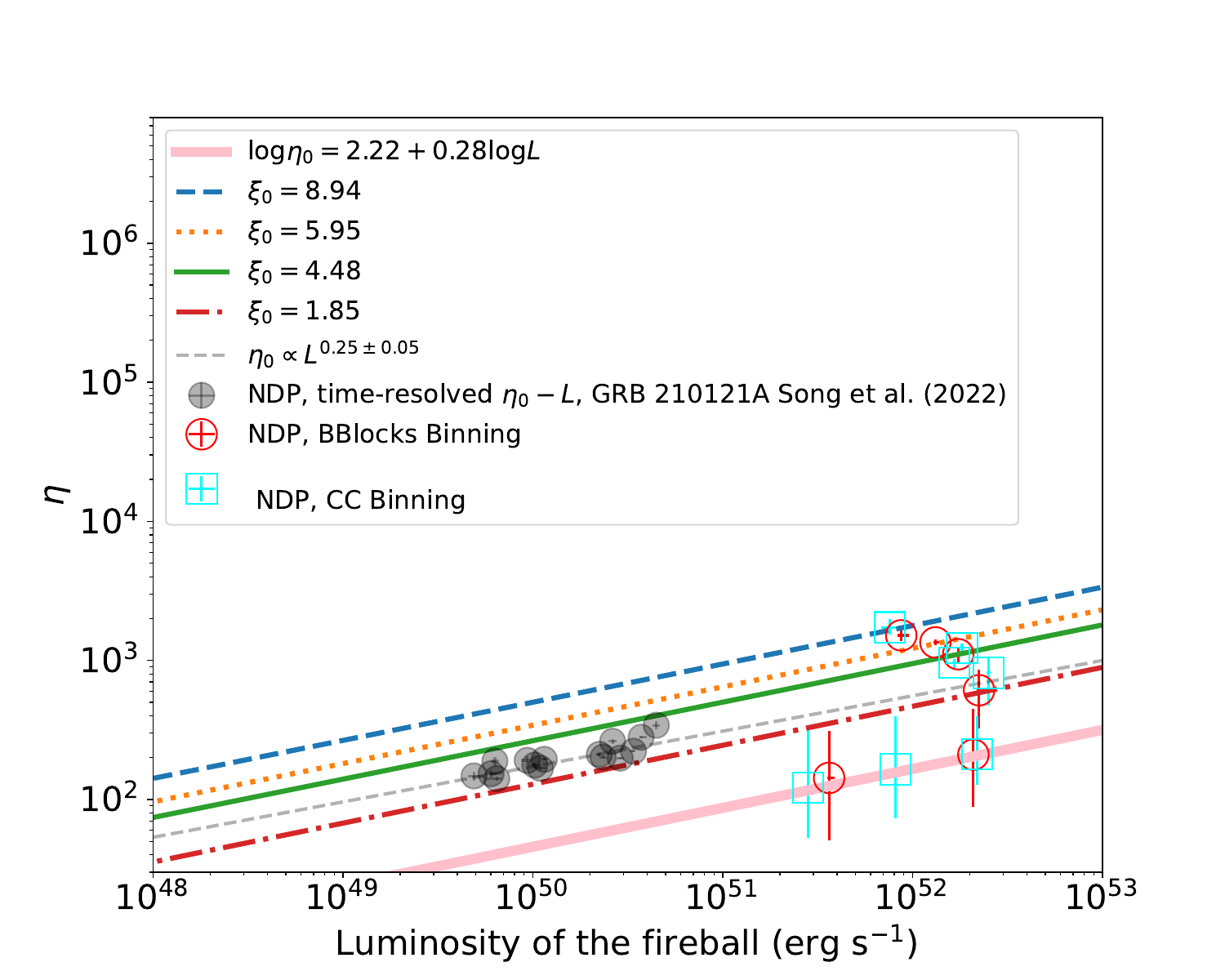}
   \put(-160,250){(c)}\\
\caption{(a) The time-integrated spectrum and the fit result with NDP+CPL in [-0.05,3.92] s. (b) The time-integrated spectrum and the fit result with mBB in [-0.05,3.92] s. (c) The correlation of $\eta-L$. The red circles denote the time-resolved $\eta-L$ estimated with BBlocks binning while the cyan squares denote that with CC binning (the same below). The dot-dashed line, the solid thin line, the
dotted line and the dashed line denote $\hat{\eta}-\hat{L}$ considering different values of neutron-proton ratio $\xi_0$. The black solid circles and thin dashed line denote $\eta-L$ of GRB 210121A. (d) $E_{\rm p,z}$ vs $E_{\rm \gamma,iso}$, adapted from Figure 7 in \protect\cite{2022arXiv220508427W}. The yellow circle denotes the position of GRB 220426A, and the same in (e). (e) $L_{\rm iso}$ versus $\tau_{RF}$, adapted from Figure 5 in \protect\cite{2015MNRAS.446.1129B}.
(f) $\Gamma $ versus $r_0$, adapted from Figure 1 in \protect\cite{2015ApJ...813..127P}. The yellow circle denotes $\eta$ versus $r_0$ of GRB 220426A, while the red one denotes $\Gamma$ versus $r_0$.
\label{fig:NDPres}
}
\end{center}
\end{figure*}

We first perform coarse fits per bin individually and find that in the first $\sim4$ s, the emission is dominated by the unsaturated regime with large $\eta$, while after 4 s, $\eta$ is much smaller. To suppress the uncertainty, $z$ should be extracted from enough statistics, and the value of $z$ is fixed in the time-resolved analysis.  Therefore, a time-integrated fit is performed in [-0.05, 3.92] s. A time-integrated fit is performed with a combined model of NDP and IC contribution described by a CPL function. To be  distinguished from the lower energy photon index $\alpha$ in BAND and CPL functions used in Section~\ref{sec:modeling}, the lower energy photon index of the spectrum of this additional component is denoted as $B$. The results is shown in Figure~\ref{fig:NDPres} (a) and listed in Table~\ref{tab:GRB220426A}. From Figure~\ref{fig:NDPres} (a) and (b),  $\Delta$BIC between NDP+CPL and mBB model is 28.0 with the changes of degree of freedom (dof) equal to 5.
The spectrum of the non-thermal component seems very narrow with a large $B=2.17^{+0.33}_{-0.15}$ and similar to thermal emission rather than synchrotron emission. Thus, it seems from the inverse Compton Scattering of photons from NDP by electrons, and the seed photons are from the thermal emissions. Note that in PHF, there are sideway diffusion effect of photons at certain
angular distances, and this sideways diffusion can cause a smearing
out effect on temperature and lead to a nonthermal spectrum
due to inverse-Compton radiation~\citep{2013ApJ...777...62I,2013A}.  Besides, its flux has an order of $\sim{10^{-7}}$ erg cm$^{-2}$ s$^{-1}$, much less than that of the average flux ($\sim 10^{-5}$ erg cm$^{-2}$ s$^{-1}$) in the first 4 s. Even if it is from other mechanisms, its contribution to the total emission is small.

\begin{figure*}
\begin{center}
 \centering      \includegraphics[width=0.6\textwidth]{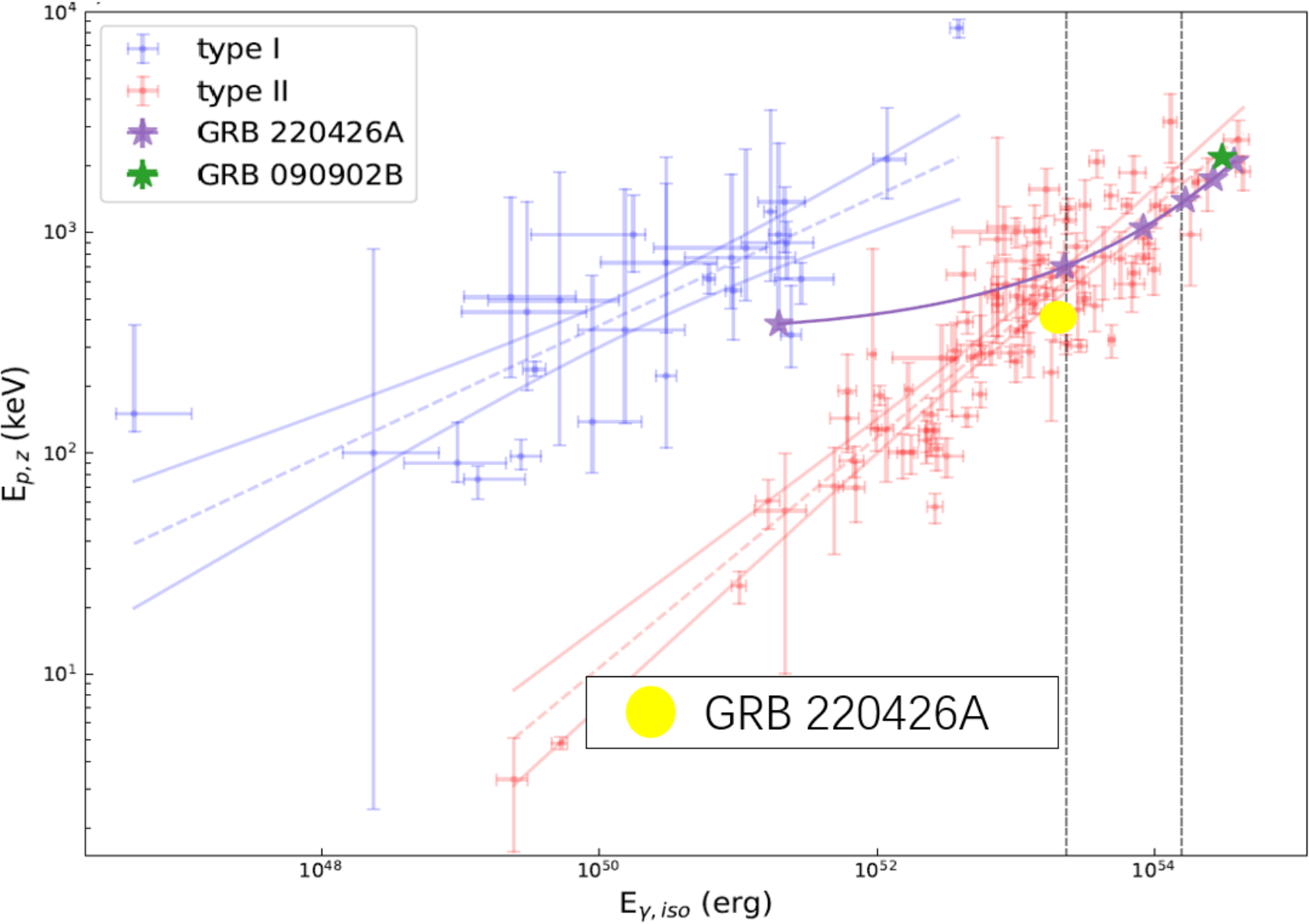}
   \put(-160,200){(d)}\\
\includegraphics[width=0.65\textwidth]{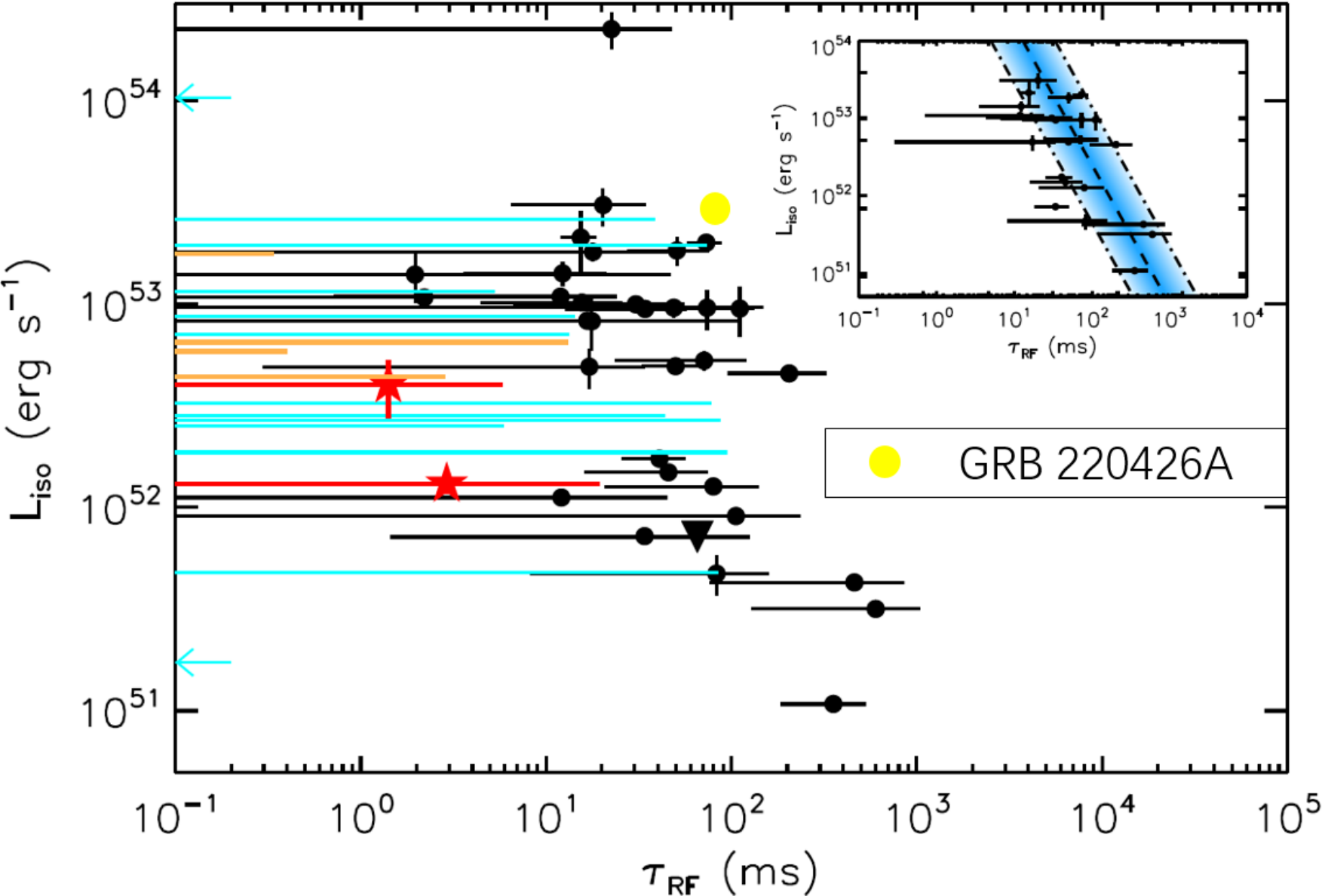}
   \put(-160,200){(e)}\\
\includegraphics[width=0.6\textwidth]{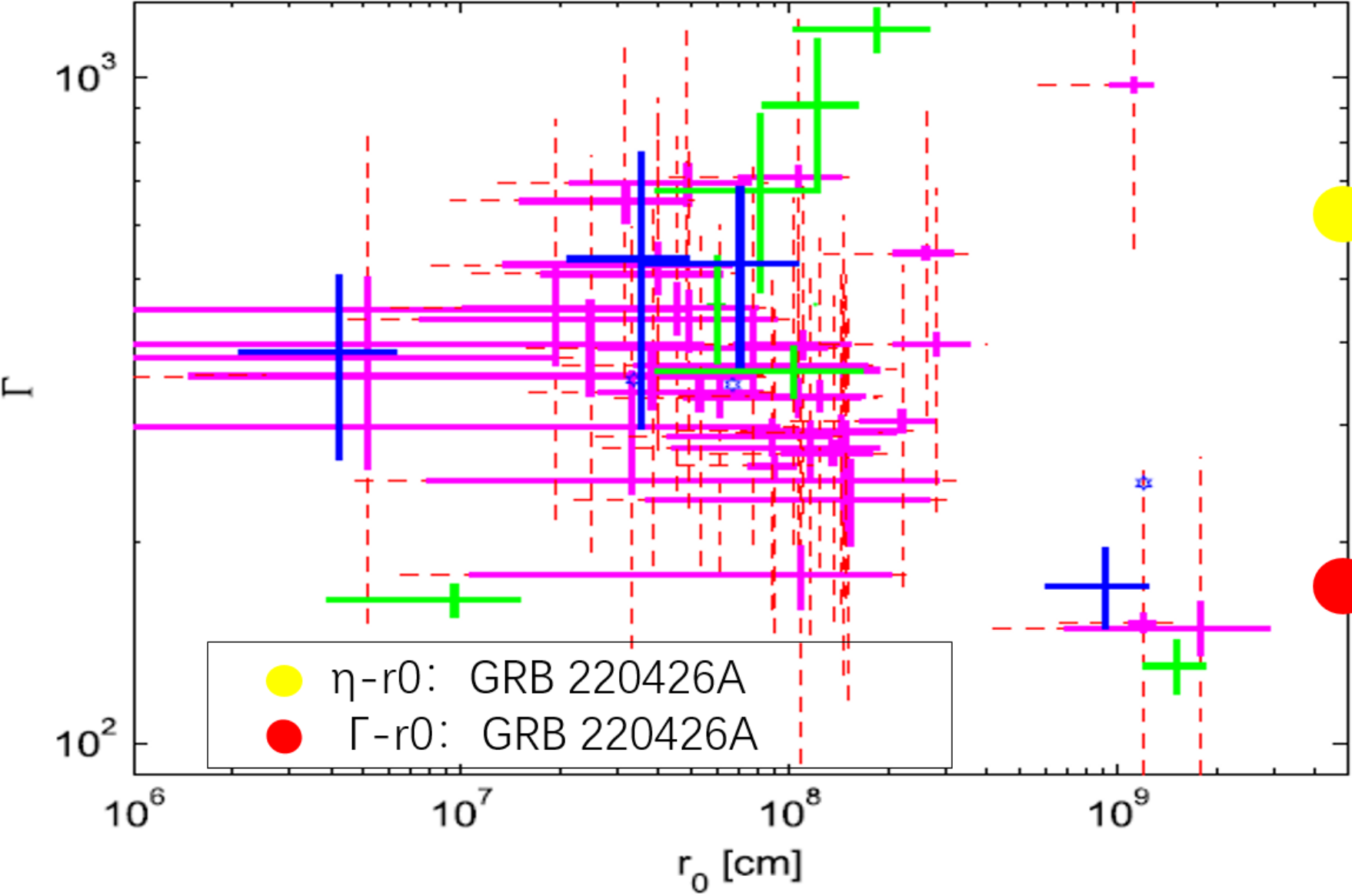}
   \put(-160,180){(f)}
\center{Figure~\ref{fig:NDPres} (Continued.)}
\end{center}
\end{figure*}

\begin{table*}
\tiny
\caption{ Fit results of time-integrated and time-resolved spectra with NDP+CPL model. The time bins in red denote CC bins.}
\begin{threeparttable}
\begin{tabular}{l|cccccccccccccc}\hline\hline
time bins &log$(r_0)$ & $\eta_0$  &p &$\theta_c$ &log$(L$) &z  
& $C$ &$B$ &$E_{\rm p}$
&BIC
&$\frac{\chi^2}{dof}$
&UNSAT & $\Gamma$ &$R_{\rm ph}$\\
(s) &(cm) & & & &(erg s$^{-1}$) & & & &(keV) & & 
& & &$\times10^{11}$ cm\\\hline

\label{tab:GRB220426A}
[-0.05,7.79] &9.54$^{+0.01}_{-0.02}$ &584.2$^{+111.6}_{-35.3}$ &12.51$^{+3.98}_{-9.21}$ &0.13$^{+0.01}_{-0.10}$ &51.89$^{+0.17}_{-0.04}$ &1.28$^{+0.18}_{-0.09}$ &5.60$^{+1.53}_{-0.32}$ &2.33$^{+0.16}_{-0.08}$ &403.24$^{+5.13}_{-49.64}$ &426.9 &$\frac{405.7}{217}$ &Y &131.0 &4.54\\\hline
[-0.05, 3.92] &9.41$^{+0.02}_{-0.02}$
 &1133.8$^{+171.2}_{-116.7}$
 &9.12$^{+5.38}_{-4.11}$
 &0.10$^{+0.03}_{-0.05}$
 &52.20$^{+0.14}_{-0.18}$
 &1.47$^{+0.24}_{-0.31}$
 &2.81$^{+1.11}_{-1.33}$
 &2.17$^{+0.33}_{-0.15}$
 &539.65$^{+66.30}_{-143.89}$
 &351.7 &$\frac{330.5}{217}$ 
 &Y &147.2 &3.78 
 \\\hline
[-0.05, 2.22] &9.32$^{+0.02}_{-0.03}$ &1346.6$^{+67.8}_{-67.2}$ &13.43$^{+4.88}_{-6.47}$ &0.12$^{+0.05}_{-0.04}$ &52.12$^{+0.01}_{-0.02}$&1.47 &2.58$^{+1.56}_{-1.36}$ &2.12$^{+0.21}_{-0.23}$ &539.93$^{+61.04}_{-95.92}$ &252.0 &$\frac{233.7}{190}$
&Y& 140.10  &2.93
\\ \hline
[2.22, 3.42] &9.42$^{+0.01}_{-0.01}$ &610.4$^{+247.6}_{-288.1}$ &12.68$^{+4.39}_{-7.05}$ &0.13$^{+0.03}_{-0.02}$ &52.35$^{+0.01}_{-0.01}$&1.47 &2.42$^{+1.87}_{-1.06}$ &2.14$^{+0.45}_{-1.13}$ &575.76$^{+61.24}_{-203.61}$ &285.0 &$\frac{267.2}{160}$
&Y & 201.52  &5.30
\\ \hline
[3.42, 3.92] &9.66$^{+0.02}_{-0.02}$ &1108.0$^{+105.6}_{-166.4}$ &12.84$^{+5.02}_{-5.69}$ &0.13$^{+0.03}_{-0.21}$ &52.24$^{+0.01}_{-0.03}$&1.47 &5.34$^{+1.07}_{-1.68}$ &1.85$^{+0.52}_{-0.46}$ &565.41$^{+274.30}_{-216.86}$ &202.1 &$\frac{184.9}{130}$
&Y & 126.28  &5.77 
\\ \hline
[3.92, 5.07] &9.74$^{+0.01}_{-0.01}$ &210.2$^{+237.2}_{-121.4}$ &9.12$^{+5.39}_{-3.00}$ &0.10$^{+0.04}_{-0.22}$ &52.32$^{+0.01}_{-0.01}$&1.47 &25.43$^{+4.87}_{-3.06}$ &2.03$^{+0.43}_{-0.82}$ &391.39$^{+200.00}_{-264.72}$ &203.7 &$\frac{186.6}{130}$
&N & 219.78  &12.10 
\\ \hline
[5.07, 7.79] &9.86$^{+0.02}_{-0.05}$ &143.0$^{+170.0}_{-92.4}$ &9.15$^{+3.00}_{-3.80}$ &0.15$^{+0.04}_{-0.05}$ &51.56$^{+0.01}_{-0.03}$&1.47 &6.79$^{+1.50}_{-2.99}$ &2.20$^{+0.11}_{-0.17}$ &304.31$^{+243.35}_{-34.45}$ &206.1 &$\frac{189.3}{120}$
&N & 127.18 &9.21 
\\ \hline\hline
{\color{red}{
[-0.05, 0.95]}} &9.13$^{+0.12}_{-0.11}$ &1726.8$^{+251.0}_{-190.3}$ &10.95$^{+10.20}_{-12.59}$ &0.11$^{+0.10}_{-0.11}$ &51.88$^{+0.03}_{-0.04}$&1.47 &2.46$^{+2.49}_{-2.24}$ &2.06$^{+0.30}_{-0.19}$ &553.60$^{+78.34}_{-121.87}$ &196.6 &$\frac{178.8}{160}$ 
&Y & 124.10 &1.67 
\\\hline
{\color{red}{[0.95, 1.95]}} &9.37$^{+0.01}_{-0.02}$ &1228.0$^{+99.8}_{-83.1}$ &13.66$^{+13.49}_{-10.19}$ &0.14$^{+0.15}_{-0.09}$ &52.26$^{+0.01}_{-0.01}$&1.47 &2.24$^{+1.82}_{-1.74}$ &1.95$^{+0.89}_{-0.58}$ &574.93$^{+253.12}_{-210.01}$ &194.6 &$\frac{177.1}{145}$ 
&Y & 154.80 &3.63 
\\\hline
{\color{red}{[1.95, 2.95]}} &9.37$^{+0.01}_{-0.01}$ &814.5$^{+230.9}_{-339.7}$ &13.05$^{+8.39}_{-10.79}$ &0.11$^{+0.10}_{-0.14}$ &52.40$^{+0.01}_{-0.01}$&1.47 &2.24$^{+1.62}_{-1.75}$ &2.08$^{+1.05}_{-0.61}$ &604.41$^{+266.54}_{-232.37}$ &208.4 &$\frac{191.4}{124}$ 
&Y & 197.64 &4.63
\\\hline
{\color{red}{[2.95, 3.95]}} &9.58$^{+0.01}_{-0.01}$ &963.2$^{+56.0}_{-84.0}$ &10.95$^{+16.20}_{-6.90}$ &0.12$^{+0.04}_{-0.08}$ &52.22$^{+0.01}_{-0.01}$&1.47 &2.36$^{+1.83}_{-1.37}$ &1.69$^{+0.66}_{-0.41}$ &594.67$^{+280.00}_{-289.94}$ &233.5 &$\frac{216.4}{130}$ 
&Y & 138.55 &5.27
\\\hline
{\color{red}{[3.95, 4.95]}} &9.74$^{+0.01}_{-0.01}$ &211.3$^{+185.3}_{-83.2}$ &12.05$^{+14.39}_{-7.80}$ &0.11$^{+0.11}_{-0.11}$ &52.34$^{+0.01}_{-0.01}$&1.47 &24.04$^{+1.85}_{-3.99}$ &1.92$^{+0.89}_{-0.32}$ &458.11$^{+263.73}_{-242.40}$ &203.1 &$\frac{186.0}{130}$ 
&N & 222.78 &12.22
\\\hline
{\color{red}{[4.95, 5.95]}} &9.82$^{+0.01}_{-0.02}$ &164.9$^{+234.5}_{-92.0}$ &13.16$^{+3.00}_{-2.38}$ &0.09$^{+0.06}_{-0.15}$ &51.91$^{+0.01}_{-0.02}$&1.47 &18.04$^{+1.00}_{-4.93}$ &1.84$^{+0.32}_{-0.82}$ &404.43$^{+276.00}_{-201.23}$ &174.4 &$\frac{157.5}{120}$ 
&N & 163.59 &10.80
\\\hline
{\color{red}{[5.95, 6.95]}} &9.89$^{+0.01}_{-0.06}$ &121.8$^{+213.1}_{-69.0}$ &13.65$^{+13.49}_{-5.40}$ &0.13$^{+0.10}_{-0.09}$ &51.45$^{+0.02}_{-0.02}$&1.47 &29.11$^{+5.06}_{-4.86}$ &1.33$^{+0.30}_{-0.44}$ &473.11$^{+226.66}_{-277.48}$ &176.3 &$\frac{159.4}{120}$ 
&N & 120.50 &9.35
\\\hline
\hline
\end{tabular}
\begin{tablenotes}
\footnotesize
\item UNSAT: satisfies $\theta_{\rm cri}>5/\eta_0$ and $\theta_{\rm c}$. Y: Yes, N: No. $\theta_{\rm cri}$ is a critical angle; the regime turns from unsaturated to saturated for $\theta>\theta_{\rm cri}$, where $\theta$ is the angle measured from the jet axis.
\end{tablenotes}
\end{threeparttable}
\end{table*}

 Note that $z=1.47^{+0.24}_{-0.31}$ determined in the time-integrated results is consistent well with that predicted by Amati-relation in some work, $z=1.40^{+1.49}_{-0.38}$~\citep[][]{2022arXiv220508427W}. It is fixed when the time-resolved analyses are performed, while the other parameters are float. The fit results of BBlocks binning are shown in Table~\ref{tab:GRB220426A} and Figure~\ref{sec:TRspectra} in Appendix. The results of CC binning are also listed in Table~\ref{tab:GRB220426A} in the last 7 rows in red. 
\begin{figure*}
\begin{center}
 \centering
   \includegraphics[width=0.75\textwidth]{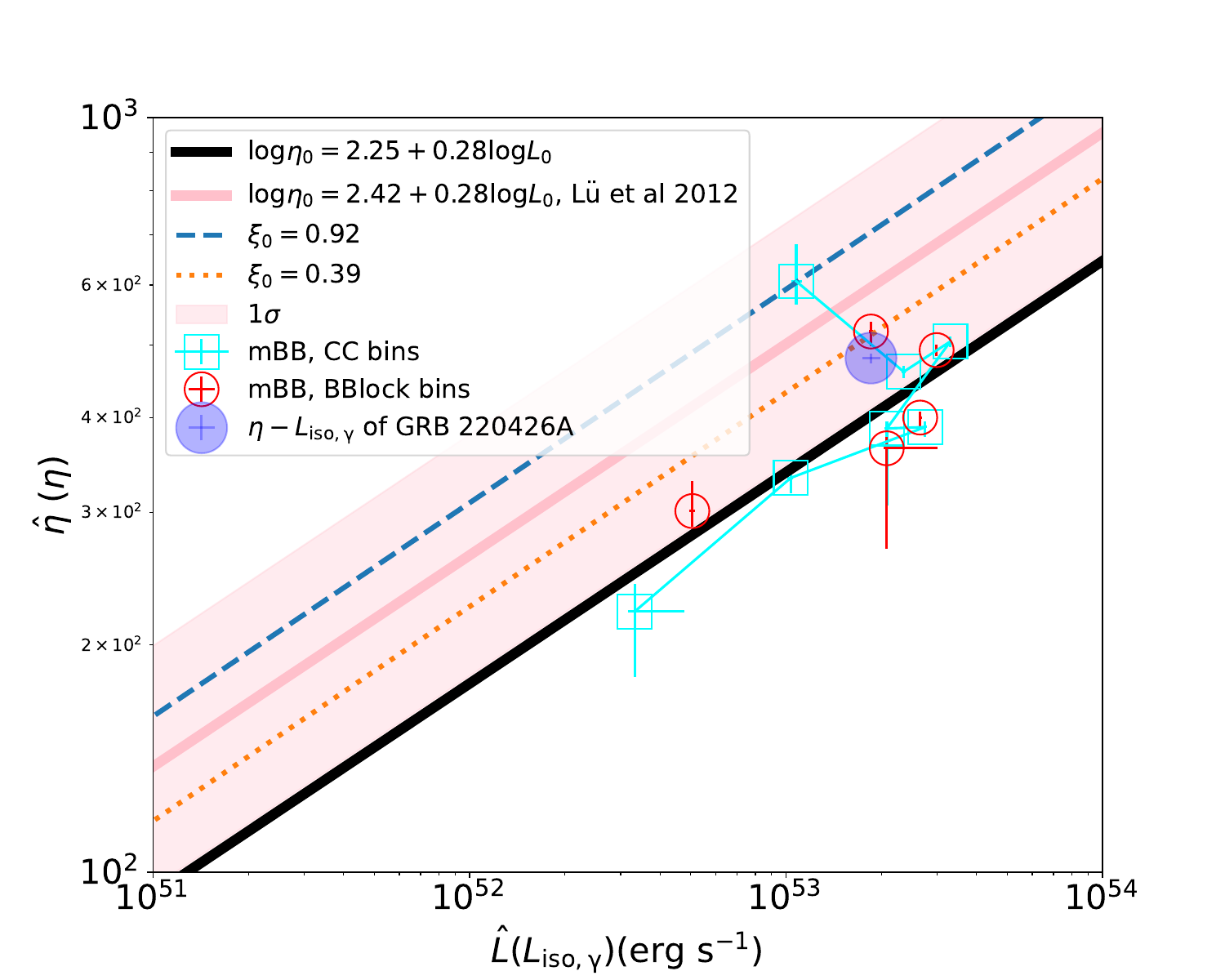}
\caption{ The measured $\eta$ versus $L_{\rm iso, \gamma}$, which is also $\hat{\eta}$ versus $\hat{L}$ after considering the n-p decoupling effect. The pink thick solid line and pink shadow represent the $\eta-L_{\rm iso, \gamma}$ and 1$\sigma$ uncertainty extracted in \protect\cite{L_2012}. The blue solid circle denotes the time-averaged $\eta$ versus $L_{\rm iso, \gamma}$ of GRB 220426A. The dashed and dotted lines denote $\hat{\eta}$ versus $\hat{L}$ with different values of $\xi_0$. 
\label{fig:etaL0} }
\end{center}
\end{figure*}

We note that in the first 1 s, the jet has a very large $\eta\sim1700$\footnote{We note that the jet has a large $p$, which is the power-law index of baryon loading parameter profile; this means that for $\theta>\theta_{\rm c}$, $\eta(\theta)\sim\frac{\eta_{0}}{
((\theta /\theta_{\rm c})^{2p}+1)^{1/2}}$ decreases sharply and can be ignored, while for $\theta<\theta_{\rm c}$, $\eta(\theta)\sim\eta_0$ ($\theta_{\rm c}$, $\eta_{0}$ and $\theta$ are defined in the introduction of NDP model in Section~\ref{sec:4funcs}). Thus, we use the value of $\eta_0$ to approximate $\eta$ of the outflow hereafter. }. However, the fireball produced by neutrino annihilation mechanism is dirty, and it is difficult to produce $\eta\sim 1700$ at $L\sim10^{52}$ erg s$^{-1}$, as shown in Figure 1 in \cite{2013ApJ...765..125L}. 
The measured $\eta-L$ is shown in Figure~\ref{fig:NDPres} (c), where the red circles denote that of BBlocks Binning, while the cyan squares denote that of CC Binning. The trends of these two binning methods are similar, which means that the trend of $\eta-L$ is not sensitive to the locations of the time bins. 

For further comparison, we choose GRB 210121A which is dominated by typical photospheric emission from a pure hot fireball~\citep{2021ApJ...922..237W, 2022arXiv220409430S}.
 $\eta-L$ of GRB 210121A 
 is also extracted by the similar method and procedure. It is plotted in black circles and the index is $0.25\pm0.05$~\citep{2022arXiv220409430S} as shown in black thin dashed line, which seems a monotonically
positive correlation and consistent well with that of $\nu\overline{\nu}$ annihilation mechanism. It seems that the correlation of $\eta-L$ of GRB 220426A is abnormal in the first few seconds. 

We also have some tests on other properties. With $E_{\rm p}$ and flux obtained from time-integrated spectral fit in [-0.05,7.79] s with the BAND function, the isotropic-equivalent radiated energy $E_{\rm iso,\gamma}\sim 2.2 \times 10^{53}$ erg with $z=1.47$. It is much closer to a type II GRB according to the empirical $E_{\rm p}(1+z)-E_{\rm iso,\gamma}$ correlation as shown in Figure~\ref{fig:NDPres} (d). Figure~\ref{fig:NDPres} (e), adapted from Figure 5 in \cite{2015MNRAS.446.1129B}, shows the correlation of the isotropic equivalent luminosity $L_{\rm iso}$ and the rest-frame lags $\tau_{\rm RF}$ between the rest frame energy bands of [100, 150] keV and [200, 250] keV. With the discrete cross-correlation function~\citep[e.g.,][]{2010ApJ...711.1073U}, $\tau_{\rm RF}$ of GRB 220426A is determined to be $81\pm2$ ms. Given $L_{\rm iso}\sim 1.8\times10^{53}$ erg s$^{-1}$, $L_{\rm iso}-\tau_{\rm RF}$ of GRB 220426A is also consistent with that of long GRBs. The model-independent $E_{\rm p}(1+z)-E_{\rm iso,\gamma}$ and $L_{\rm iso}-\tau_{\rm RF}$ of GRB 220426A are consistent well with those of long GRBs, which means that its jet launching mechanism is not different from those of most long GRBs.

In~\cite{2015ApJ...813..127P}, 47 GRBs with significant thermal emission components are identified, and $10^2<\Gamma<10^3$ with an average $\langle\Gamma\rangle=370$, and $10^{6.5}<r_0<10^{9.5}$ cm with $\langle r_0\rangle=10^8$ cm are reported.
Although the averaged $\Gamma\sim130$ (shown in the first row in Table~\ref{tab:GRB220426A}, note that the regime determined by NDP model is unsaturated as shown in the 13th column in Table~\ref{tab:GRB220426A}) and the averaged $r_0\sim 10^{9.5}$ cm in GRB 220426A are not substantially outside, but are on the edge of the previously observed distribution, as shown in Figure.~\ref{fig:NDPres} (f).
 
 In summary, the extremely large $\eta$ in the initial time of the burst requires a clean outflow in the early stage. It seems not usual in a pure fireball produced in $\nu\overline{\nu}$ annihilation mechanism. $\eta-L$ seems abnormal compared with that of GRB 210121A which is from a hot pure fireball. The fireball property, $\Gamma-r_0$ is not typical as well. Thus, this may be not a possible source, or it may be because NDP model or the jet structure used here is not proper.  In the next section, the other model, mBB, is used to extract $\eta$ and $\eta-L$.
 
 \subsection{B. A hybrid outflow with completely thermalized magnetic dissipations}\label{sec:b.}
 We assume the simplest case that the magnetic energy is completely thermalized below the photosphere and converted to heat. Note that it could be achieved because of the existence of the extra thermal component in the outflow. In this case, the dimensionless entropy $\mu_0=\eta_{\rm m}(1+\sigma_0)=\frac{L_{\rm m}+L_{\rm p}}{\dot{M}c^2}$, where $\sigma_0$ is the ratio of Poynting flux luminosity to the matter flux; $\dot{M}$ is the mass rate of baryon loading; $L_{\rm p}$ is the luminosity of Poynting flux and $L_{\rm m}$ is the luminosity of  matter flux, and $\eta_{\rm m}=\frac{L_{\rm m}}{\dot{M}c^2}$. 
 We investigate $\eta$ and time-resolved $\eta-L$ with mBB model. \cite{2007ApJ...664L...1P,2015ApJ...813..127P} introduce a method as
 \begin{equation}
		\Gamma=[(1.06)(1+z)^{2}d_{\rm L}\frac{Y\sigma_{\rm T}F^{\rm ob}}{2m_{p}c^{3}\mathcal{R}}]^{1/4},
		\label{Lorentz}
	 \end{equation}
	where $d_{\rm L}$ is the luminosity distance, $\sigma_{\rm T}$ is the Thomson scattering cross section, and $F^{\rm ob}$ is the observed flux. $Y$ is the ratio between the total outflow energy and the energy emitted in the gamma rays, and $Y\geq1$. Because of non-detection of afterglow, $Y$ is taken to be 1 in our analysis. $\mathcal{R}=(\frac{F^{\rm ob}_{\rm thermal}}{\sigma T^{4}_{\rm max}})^{1/2}$ where $F^{\rm ob}_{\rm thermal}$ is the thermal emission flux. $\sigma$ is Stefan–Boltzmann constant. Note that in this method, the emission is assumed to be from the saturated regime, thus $\Gamma$ reaches the value of $\eta$. We use $\eta$ in the following statement. In this case, $r_0$ is determined to be $0.6\frac{d_{\rm L}}{(1+z)^2}(\frac{F^{\rm ob}_{\rm thermal}}{YF_{\rm ob}})^{3/2}\mathcal{R}$ with a unit of cm.
	
	The fit results are shown in Table~\ref{tab:TRmBB} and Figure~\ref{fig:TRres_floatpara2}\footnote{ Note that in some time bins, there exists small inconsistency in a few energy bins in high energy band. However, even if an additional component in high energy band is added in the fit, the extracted parameters are consistent well with those only with mBB, with changes $\lesssim5\%$. The contribution from so-called other mechanism is very small. Thus we think it is not necessary to fit with mBB plus on an additional component.}. The isotropic $\gamma$-ray luminosity $L_{\rm iso, \gamma}$ is determined with the observed flux and $z$. The trend of $\eta-L_{\rm iso, \gamma}$ is shown in Figure~\ref{fig:etaL0}. The time-averaged $\eta-L_{\rm iso, \gamma}$ denoted by blue solid circle is well consistent with L{\"u}'s relation. Note that there is a relation between $L$ and $L_{\rm iso, \gamma}$, $L=f_{\rm b}L_{\rm iso, \gamma}$, where $f_{\rm b}\ll1$ is the beaming factor. It is reasonable to assume that $f_{\rm b}$ could be taken as a constant during the burst. Thus, the time-resolved trend of $\eta-L$ is the same as that of $\eta-L_{\rm iso, \gamma}$. The values of $\eta$ estimated with mBB model are smaller than those with NDP model, and this may be because the regime determined from fit results with NDP model is unsaturated while it is assumed to be saturated with the method in \cite{2007ApJ...664L...1P,2015ApJ...813..127P}. However, it seems interesting that the trend of $\eta-L$ obtained with mBB model is similar to that of NDP model and like a symbol of `>'. Thus, the outflow in GRB 220426A has larger $\eta$ with a lower luminosity in the first few seconds, which may be model-independent to some extent. 
\begin{table*}
\caption{ Time-resolved results of fit with mBB model. The time bins in red denote CC bins.}\label{tab:TRmBB}
\begin{tabular}{l|cccccccc}\hline\hline
Time bins(s) &$m$ &$kT_{\rm min}$(keV)  &$kT_{\rm max}$(keV) &flux (10$^{-6}$ erg cm$^{-2}$ s$^{-1}$)  &BIC &$\frac{\chi^2}{ndof}$ &$\eta$ &log$r_0$(cm) \\\hline
[-0.05, 7.79]& -0.39$^{+0.07}_{-0.08}$ &11.47$^{+0.44}_{-0.46}$ &95.10$^{+3.00}_{-2.63}$ &12.96$^{+0.09}_{-0.08}$ &459.6 &$\frac{450.2}{222}$ &480.09$^{+7.61}_{-6.84}$ &9.11$^{+0.02}_{-0.03}$\\\hline
[-0.05, 2.22]& -0.44$^{+0.19}_{-0.27}$ &19.38$^{+1.25}_{-1.87}$ &110.89$^{+9.96}_{-7.22}$ &13.0$^{+0.2}_{-0.2}$ &214.5 &$\frac{205.6}{164}$ &520.6$^{+23.3}_{-15.1}$ &8.99$^{+0.05}_{-0.05}$\\\hline
[2.22, 3.42]& 0.31$^{+0.12}_{-0.15}$ &11.08$^{+1.33}_{-1.06}$ &89.42$^{+3.17}_{-2.69}$ &21.0$^{+0.2}_{-0.2}$ &282.1 &$\frac{273.4}{149}$ &491.9$^{+9.0}_{-7.7}$ &9.28$^{+0.03}_{-0.03}$\\\hline
[3.42, 3.92]& 0.81$^{+0.00}_{-0.00}$ &10.01$^{+2.00}_{-4.00}$ &53.49$^{+2.60}_{-2.77}$ &14.6$^{+0.4}_{-0.2}$ &202.7 &$\frac{194.3}{128}$ &364.6$^{+96.4}_{-13.2}$ &9.66$^{+0.05}_{-0.10}$\\\hline
[3.92, 5.07]& 0.07$^{+0.19}_{-0.21}$ &9.01$^{+1.97}_{-1.11}$ &59.82$^{+2.81}_{-2.15}$ &18.6$^{+0.2}_{-0.2}$ &233.0 &$\frac{224.4}{134}$ &400.4$^{+10.1}_{-7.3}$ &9.61$^{+0.03}_{-0.04}$\\\hline
[5.07, 7.79]& -1.38$^{+0.43}_{-0.40}$ &9.66$^{+0.71}_{-0.93}$ &57.43$^{+19.78}_{-12.77}$ &3.5$^{+0.1}_{-0.1}$ &234.8 &$\frac{226.3}{134}$ &301.0$^{+14.3}_{-28.6}$ &9.37$^{+0.23}_{-0.11}$\\\hline\hline
{\color{red}{[-0.05, 0.95]}}& -1.50$^{+0.67}_{-0.63}$ &33.75$^{+3.44}_{-3.85}$ &173.68$^{+25.85}_{-38.78}$ &7.5$^{+0.3}_{-0.3}$ &193.5 &$\frac{184.6}{164}$ &606.9$^{+42.0}_{-72.1}$ &8.47$^{+0.25}_{-0.19}$\\\hline
{\color{red}{[0.95, 1.95]}}& 0.46$^{+0.26}_{-0.29}$ &14.14$^{+1.93}_{-1.69}$ &81.70$^{+4.28}_{-4.82}$ &16.5$^{+0.2}_{-0.2}$ &197.2 &$\frac{188.5}{149}$ &460.5$^{+8.3}_{-8.3}$ &9.29$^{+0.02}_{-0.04}$\\\hline
{\color{red}{[1.95, 2.95]}}& 0.49$^{+0.15}_{-0.16}$ &11.15$^{+1.73}_{-1.38}$ &88.88$^{+3.48}_{-4.81}$ &23.1$^{+0.2}_{-0.3}$ &241.3 &$\frac{232.8}{128}$ &505.1$^{+8.4}_{-8.3}$ &9.29$^{+0.04}_{-0.02}$\\\hline
{\color{red}{[2.95, 3.95]}}& 0.52$^{+0.35}_{-0.38}$ &11.94$^{+1.99}_{-2.26}$ &60.22$^{+2.39}_{-2.49}$ &14.6$^{+0.2}_{-1.2}$ &240.2 &$\frac{231.7}{134}$ &387.3$^{+81.4}_{-8.3}$ &9.55$^{+0.04}_{-0.10}$\\\hline
{\color{red}{[3.95, 4.95]}}& 0.29$^{+0.23}_{-0.23}$ &8.11$^{+1.45}_{-1.62}$ &57.84$^{+3.27}_{-2.11}$ &19.3$^{+0.3}_{-0.4}$ &235.8 &$\frac{227.2}{134}$ &389.0$^{+11.1}_{-7.3}$ &9.66$^{+0.05}_{-0.30}$\\\hline
{\color{red}{[4.95, 5.95]}}& -0.81$^{+0.39}_{-0.50}$ &9.68$^{+3.60}_{-1.13}$ &58.45$^{+12.56}_{-5.27}$ &7.2$^{+0.2}_{-0.3}$ &165.9 &$\frac{157.5}{124}$ &333.1$^{+15.6}_{-15.6}$ &9.52$^{+0.06}_{-0.06}$\\\hline
{\color{red}{[5.95, 6.95]}}& -0.05$^{+0.82}_{-0.96}$ &7.01$^{+3.87}_{-3.87}$ &30.69$^{+5.00}_{-5.10}$ &2.3$^{+0.1}_{-1.4}$ &173.0 &$\frac{164.6}{124}$ &221.5$^{+40.0}_{-19.4}$ &9.72$^{+0.15}_{-0.10}$\\\hline

\end{tabular}
\end{table*}
 
Note that we use the smallest $Y=1$ in estimation of $\eta$, and $\eta$ could be larger if $Y>1$ is used.
The maximum $\eta$ reaches up to $600$ in the first 1 s, and it is larger than the mean value (370) of the previous distribution of observed $\eta$. It is still large for $\nu\overline{\nu}$ annihilation mechanism because there is too much baryon contamination~\citep{2013ApJ...765..125L}. Moreover, we do not have found any natural explanation for the trend of time-resolved $\eta-L$ even if we think the large $\eta$ could be produced by $\nu\overline{\nu}$ annihilation mechanism.

Besides of $\nu\overline{\nu}$ annihilation, the other jet launching mechanism is the Blandford \& Znajek mechanism~\cite[BZ,][]{1977MNRAS.179..433B}, in which the spin energy of the BH is tapped by a magnetic field, and produces a Poynting flux. Due to the magnetic field, the outflow is cleaner~\citep[e.g.,][]{2013ApJ...765..125L}. As discussed in ~\cite{2013ApJ...765..125L} and \cite{YI20171}, the neutron drift rate into the jet is $\dot{M}_{n}\propto \dot{m}^{0.83}$ and the power $L_{\rm B}\propto \dot{m}$. 
 The index of $\mu_0-L$ could be $\sim0.17$ or $\sim 0.30$ with or without adding an extra $\nu\overline{\nu}$ annihilation power, thus in case of hybrid outflow, the index$\sim 0.30$.
If there is a natural scenario which could change the baryon loading $\dot{M}_{n}$, it may account for the measured trend of time-resolved $\eta-L$.

 We note that in the case of BZ mechanism, because of the existence of magnetic fields, only the protons whose motions are almost aligned with the magnetic field lines from the field line footing on the disk, can be ejected into the atmosphere, whereas neutrons can penetrate
magnetic field lines and drift from sideways into the jet during the propagation, or from the disk. In the outflow, the electrons can exchange energy with photons via Compton scattering and with protons via Coulomb collisions; however, for the neutrons, only protons have enough mass to collider with them and cause a collisional neutron-proton (n-p) coupling. This means that, if the neutron-to-proton ratio (denoted as $\xi_0$) is large so that there are not enough protons, or the Lorentz factor of the outflow is large enough (above the the critical value $\eta^*$ of n-p decoupling), the neutrons can not be accelerated so fast along with the other outflow component, even if they have drifted into the outflow in the early stage of acceleration. In this scenario the n-p decoupling happens~\citep{2006MNRAS.369.1797R}, and thus $\dot{M}_{n}$ is changed. Note that the n-p decoupling is difficult in pure matter flux of pure hot fireball, because there is not magnetic field to prevent protons enter the outflow.

$\hat{\eta}$ and $\hat{L}$ are defined as the Lorentz factor and luminosity after neutrons decouple from the outflow, which becomes `neutron free'. From the Equations (57), (58) and (59) in \cite{2006MNRAS.369.1797R}, $\hat{\eta}$ is calculated as
\begin{equation}\label{eq:hateta}
      \hat{\eta}=\eta+\xi_0(\eta-\frac{\eta^{*4/3}}{\eta^{1/3}}\frac{1}{1+\xi_0}),
\end{equation}\\
  and here $\eta$ denotes the baryon loading parameter before n-p decoupling, and $\eta^*\approx 4.8\times10^2(\frac{L_{52}}{(1+\xi_0)r_{0,7}})^{1/4}$ (from Equation (55) in \cite{2006MNRAS.369.1797R}). Note that the increasement $\Delta\hat{\eta}=\hat{\eta}-\eta=\xi_0(\eta- \eta^{*4/3}/\eta^{1/3}\frac{1}{1+\xi_0})$ must be larger than 0 because $\eta\geq\eta^*$. 
  $\hat{L}$ must be less than the initial luminosity $L$ before n-p decoupling, due to the loss of neutrons, and could be described as
  \begin{equation}\label{eq:hatL}
      \hat{L}=L- \frac{\eta^{*4/3}}{\eta^{1/3}}\dot{M}c^2 \frac{\xi_0}{1+\xi_0},
  \end{equation}
where $\dot{M}c^2=L/\eta$. Thus, a lower luminosity  with a larger $\hat{\eta}$ could cause a bias on the $\eta-L$ correlation.

We assume an origin correlation between $\eta$ and $L$, e.g., $\eta=a + b\log(L)$ before n-p decoupling, and for this correlation $\xi_0$ is taken to be 0. According to Equation~(\ref{eq:hateta}) and (\ref{eq:hatL}), it is taken as there is not n-p decoupling effect. Note that $\xi_0$ is a relative value rather than true value of neutron-to-proton ratio, and it could tell the changes of neutron richness during the burst. There are two unknown quantities, $L$ and $\xi_0$ in equations~(\ref{eq:hateta}) and equation~(\ref{eq:hatL}). Thus, a list of $\xi_0$ could be determined with a list of measured $\hat{\eta}$, $\hat{L}$, and $r_0$. From the fit results with NDP model, we get the decreasing values of $\xi_0$ with time: $\xi_0=$8.94, 5.95, 4.48, 1.85 for [-0.05,0.95] s, [2.22, 3.42]s, [3.42, 3.92] s, [3.92,5.07] s and $\sim0$ for the following of the burst, with $a=2.22$ and $b=0.28$ for the origin correlation. For those of mBB, $\eta=2.25 + 0.28\log(L)$ is  taken to be the origin correlation. $\xi_0=0.92$ and $\xi_0=0.39$ for the first 1 and 2 s, which also shows that $\xi_0$ decreases with time. $\hat{\eta}-\hat{L}$ with the corresponding values of $\xi_0$ is shown in Figure~\ref{fig:NDPres} (c) and Figure~\ref{fig:etaL0}.

With decreasing $\xi_0$, we speculate that the neutron richness varies from high to low during the burst, which means that the proton density increases and n-p coupling effect becomes significant with time. We also note that the estimated $r_0$ increases with time with both models as shown in Table~\ref{tab:GRB220426A} with NDP  model and Table~\ref{tab:TRmBB} with mBB model. There are two mechanisms that could cause the proton density to increase. One is that neutrons convert to protons through positron capture or n-p inelastic collision~\citep{2003ApJ...594L..19L}. For positron capture, the capture time is sensitive to $r_0$~\citep{2013ApJ...765..125L}; however, $r_0\sim10^9$ cm in GRB 220426A is two orders of magnitude larger than the typical $r_0\sim10^7$ cm, so that the positron capture may be not efficient as that in other GRBs. For large $\xi_0$ at the initial time, the density of proton is low, so that n-p inelastic collision is also inefficient. Moreover, this mechanism of neutrons converting to protons happens all the time, and can not account for the changes with time or become more significant with $r_0$. 
 The other mechanism is that the magnetic field strength decreases with $r_0$ so that the proton density is low at the beginning; as magnetic field strength decreases, protons could enter the jet and n-p coupling effect becomes significant. This provides further support for existence of magnetic field and Poynting flux in the outflow.

In summary, the existence of magnetic field could affect proton density and n-p coupling effect, so that it offer an explanation for time-resolved trend of $\eta-L$. It seems that the hybrid outflow with Poynting flux completely thermalized could account for the large $\eta$ and quasi-thermal spectrum. Thus, we think this may be a better explanation than that of pure hot fireball.

 \subsection{C. A hybrid outflow with non-dissipative photospheric emission}\label{sec:hypoD}
\begin{figure*}
\begin{center}
 \centering
   \includegraphics[width=0.65\textwidth]{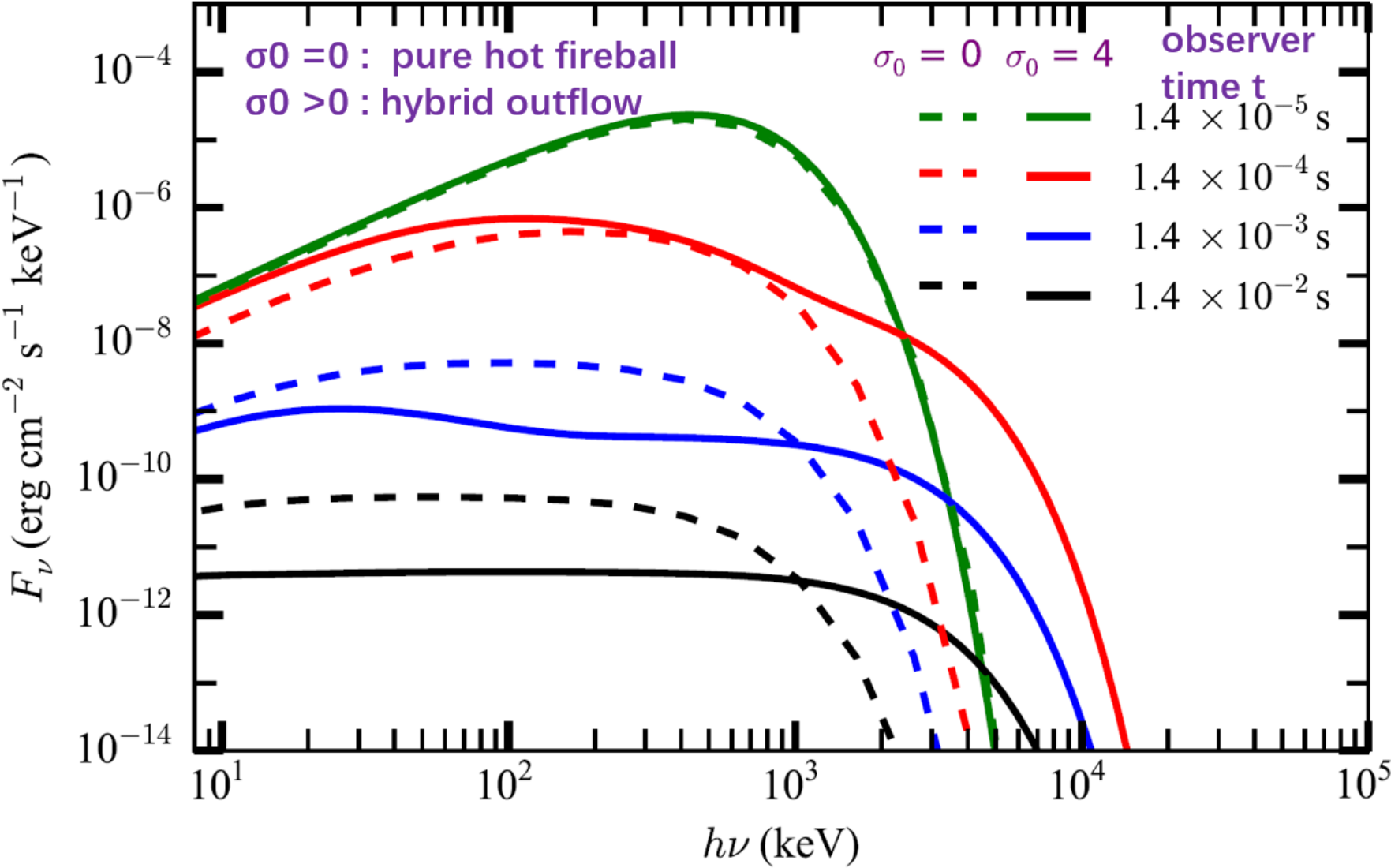}
\caption{Spectra of impulsive injection for the jet with and without magnetization adapted from Figure 1 in \protect\cite{2022MNRAS.509.6047M}. $\sigma_0$=0 and 4 represent pure fireball and hybrid outflow respectively, while the other parameters are the same. 
\label{fig:NDPMspectra}}
\end{center}
\end{figure*}

 In the non-dissipative case, all of the magnetic energy of Poynting flux is used to accelerate the outflow, which also modifies the spectrum, as shown in time-resolved spectra of an impulsive injection in Figure~\ref{fig:NDPMspectra}. The dashed lines denote spectra of NDP in PHF, while the solid lines denote those of hybrid flow with $\sigma_0=4$. At the later times (the high-latitude emission dominates), the time-resolved spectra show a power-law shape extending to a much higher energy than the early-time blackbody, and a larger $\beta$ is produced. The observed spectrum of a continuous wind is the integration of those of impulsive injections. Thus, the high-energy spectrum from such a non-dissipative hybrid outflow is a power law rather than an exponential cutoff~\citep{2022MNRAS.509.6047M}. Note that from $\alpha$ is slightly smaller in hybrid outflow from Figure~\ref{fig:NDPMspectra}, so that the spectrum still behaves like a quasi-thermal emission.
From the above discussion in Section~\ref{sec:modeling}, the spectrum of GRB 220426A has a larger $\beta$ than NDP in PHF, therefore, it seems that this hypothesis could account for the spectrum. Furthermore, considering the magnetic acceleration, the dimensionless entropy extracted in the fake hypothesis of PHF may be incorrect, thus distorts the measured trend of time-resolved $\eta-L$.
 Compared with pure hot fireball, the regimes of the photospheric emission of hybrid outflow are much more complex because the thermally driven and the magnetically driven acceleration both work~\citep{2015ApJ...801..103G}. It is very difficult to quantatively extract the properties of the jet through fit with NDP model of hybrid outflow, especially with considering a structure jet and probability emission.

However, $\epsilon_{\gamma}$ is low ($\sim10\%$) in this hypothesis~\citep{2022MNRAS.509.6047M}. This is understandable that all of the magnetic energy is transferred to kinetic energy of the outflow, which is released in afterglow phase. Thus, it is not supported, unless there exists some reason for non-detection of afterglow.
 
 \subsection{D. A hybrid outflow with not completely thermalized magnetic dissipations}\label{sec:hypoC}
 The magnetic dissipation is complex. The dissipated energy could lead to fast particles, then the synchrotron or synchrotron self Compton mechanism appears as a natural one for the prompt emission. Magnetic energy can also be dissipated directly through reconnection in a flow where the magnetic field changes polarity on small scales~\citep[e.g.,][]{2002A&A...391.1141D,2002A&A...387..714D,2006A&A...457..763G}, and produce a hot photosphere. The radius of dissipation could affect the shape of the spectrum. According to~\cite{2006A&A...457..763G}, in the saturated regime, the spectrum has a quasi-thermal appearance.
 However, luminosity of this photospheric emission is only a small part ($3\sim20\%$) of the total energy. 
 Thus, similar to hypothesis \textbf{C.}, we do not think the hot photosphere of GRB 20426A is produced by magnetic reconnection. 
 
From the simulation as shown in Figure 3 in~\cite{2015ApJ...801..103G} in this hypothesis, the spectrum could be a combination of quasi-thermal and non-thermal component. The spectrum of the former has been discussed in hypothesis \textbf{A., B.,} and \textbf{C.}. The latter could be caused by e.g. internal shocks. The energy peak of the latter is always larger than the former, which may make the total spectrum hump-like. There is no evident extra peak structure in time-integrated or time-resolved spectra (see Figure~\ref{fig:TRres_floatpara}) in higher energy band. Thus, we speculate the contribution of the dissipation is small even if it exists.

\section{summary and conclusion}\label{sec:conclusion}
 GRB 220426A is dominated by the photospheric emission. NDP model in the case of pure hot fireball as well as mBB function are used to obtain the properties of the outflow. In this analysis, we perform several tests to speculate the origin of this photospheric emission. In the hypothesis of pure hot fireball with an additional IC component in the higher energy band, the obtained $\eta$ is extremely large, which seems unreasonable for $\nu\overline{\nu}$ annihilation mechanism that produces a dirty outflow. Besides, we find the outflow has larger $\eta$ with lower luminosity in the first few seconds, from both of the fit results with NDP+IC and mBB models. A hybrid outflow with almost completely thermalized Poynting flux could account for the quasi-thermal spectra and large $\eta$. Moreover, the magnetic field could cause a varying n-p decoupling effect and account for the trend of time-resolved $\eta-L$. Besides, other origins of the photospheric emission, such as non-dissipative hybrid relativistic outflow, or magnetic reconnection, are not supported because of non-detection of afterglow. In conclusion, we think a hybrid relativistic outflow is the most likely origin. Note that a slight dissipation, or a small contribution from other mechanisms is not excluded in this case.
 
 
\section*{Acknowledgements}
The authors thank supports from the National Program on Key Research and Development Project (2021YFA0718500). This work was partially supported by International Partnership Program of Chinese Academy of Sciences (Grant No.113111KYSB20190020). The authors greatly appreciate Prof. Zi-Gao Dai for his useful comments and suggestions. The authors are very grateful to the public GRB data of Fermi/GBM. We are very grateful for the comments and suggestions of the anonymous referees. 
 Xin-Ying Song thanks Dr Jin-Zhou Wang for offering the High Definition Multimedia Interface cable. 

\section*{Data Availability}
The data underlying this article will be shared on reasonable request to the corresponding author.



\bibliographystyle{mnras}
\bibliography{GRB220426A} 

\begin{thebibliography}{}
\makeatletter
\relax
\def\mn@urlcharsother{\let\do\@makeother \do\$\do\&\do\#\do\^\do\_\do\%\do\~}
\def\mn@doi{\begingroup\mn@urlcharsother \@ifnextchar [ {\mn@doi@}
  {\mn@doi@[]}}
\def\mn@doi@[#1]#2{\def\@tempa{#1}\ifx\@tempa\@empty \href
  {http://dx.doi.org/#2} {doi:#2}\else \href {http://dx.doi.org/#2} {#1}\fi
  \endgroup}
\def\mn@eprint#1#2{\mn@eprint@#1:#2::\@nil}
\def\mn@eprint@arXiv#1{\href {http://arxiv.org/abs/#1} {{\tt arXiv:#1}}}
\def\mn@eprint@dblp#1{\href {http://dblp.uni-trier.de/rec/bibtex/#1.xml}
  {dblp:#1}}
\def\mn@eprint@#1:#2:#3:#4\@nil{\def\@tempa {#1}\def\@tempb {#2}\def\@tempc
  {#3}\ifx \@tempc \@empty \let \@tempc \@tempb \let \@tempb \@tempa \fi \ifx
  \@tempb \@empty \def\@tempb {arXiv}\fi \@ifundefined
  {mn@eprint@\@tempb}{\@tempb:\@tempc}{\expandafter \expandafter \csname
  mn@eprint@\@tempb\endcsname \expandafter{\@tempc}}}

\bibitem[\protect\citeauthoryear{{Abdo} et~al.,}{{Abdo}
  et~al.}{2009}]{2009ApJ...706L.138A}
{Abdo} A.~A.,  et~al., 2009, \mn@doi [\apjl] {10.1088/0004-637X/706/1/L138},
  \href {https://ui.adsabs.harvard.edu/abs/2009ApJ...706L.138A} {706, L138}

\bibitem[\protect\citeauthoryear{{Amati} et~al.,}{{Amati}
  et~al.}{2002}]{2002A&A...390...81A}
{Amati} L.,  et~al., 2002, \mn@doi [\aap] {10.1051/0004-6361:20020722}, \href
  {https://ui.adsabs.harvard.edu/abs/2002A&A...390...81A} {390, 81}

\bibitem[\protect\citeauthoryear{{Beloborodov}}{{Beloborodov}}{2010}]{2010Collisional}
{Beloborodov} A.~M.,  2010, \mn@doi [\mnras]
  {10.1111/j.1365-2966.2010.16770.x}, \href
  {https://ui.adsabs.harvard.edu/abs/2010MNRAS.407.1033B} {407, 1033}

\bibitem[\protect\citeauthoryear{{Bernardini} et~al.,}{{Bernardini}
  et~al.}{2015}]{2015MNRAS.446.1129B}
{Bernardini} M.~G.,  et~al., 2015, \mn@doi [\mnras] {10.1093/mnras/stu2153},
  \href {https://ui.adsabs.harvard.edu/abs/2015MNRAS.446.1129B} {446, 1129}

\bibitem[\protect\citeauthoryear{{Blandford} \& {Znajek}}{{Blandford} \&
  {Znajek}}{1977}]{1977MNRAS.179..433B}
{Blandford} R.~D.,  {Znajek} R.~L.,  1977, \mn@doi [\mnras]
  {10.1093/mnras/179.3.433}, \href
  {https://ui.adsabs.harvard.edu/abs/1977MNRAS.179..433B} {179, 433}

\bibitem[\protect\citeauthoryear{{Bromberg}, {Tchekhovskoy}, {Gottlieb},
  {Nakar}  \& {Piran}}{{Bromberg} et~al.}{2018}]{2018MNRAS.475.2971B}
{Bromberg} O.,  {Tchekhovskoy} A.,  {Gottlieb} O.,  {Nakar} E.,   {Piran} T.,
  2018, \mn@doi [\mnras] {10.1093/mnras/stx3316}, \href
  {https://ui.adsabs.harvard.edu/abs/2018MNRAS.475.2971B} {475, 2971}

\bibitem[\protect\citeauthoryear{{Burgess}}{{Burgess}}{2014}]{2014On}
{Burgess} J.~M.,  2014, \mn@doi [\mnras] {10.1093/mnras/stu1925}, \href
  {https://ui.adsabs.harvard.edu/abs/2014MNRAS.445.2589B} {445, 2589}

\bibitem[\protect\citeauthoryear{{Burrows} et~al.,}{{Burrows}
  et~al.}{2005}]{2005SSRv..120..165B}
{Burrows} D.~N.,  et~al., 2005, \mn@doi [\ssr] {10.1007/s11214-005-5097-2},
  \href {https://ui.adsabs.harvard.edu/abs/2005SSRv..120..165B} {120, 165}

\bibitem[\protect\citeauthoryear{{Cavallo} \& {Rees}}{{Cavallo} \&
  {Rees}}{1978}]{1978MNRAS.183..359C}
{Cavallo} G.,  {Rees} M.~J.,  1978, \mn@doi [\mnras] {10.1093/mnras/183.3.359},
  \href {https://ui.adsabs.harvard.edu/abs/1978MNRAS.183..359C} {183, 359}

\bibitem[\protect\citeauthoryear{{Chen} \& {Beloborodov}}{{Chen} \&
  {Beloborodov}}{2007}]{2007ApJ...657..383C}
{Chen} W.-X.,  {Beloborodov} A.~M.,  2007, \mn@doi [\apj] {10.1086/508923},
  \href {https://ui.adsabs.harvard.edu/abs/2007ApJ...657..383C} {657, 383}

\bibitem[\protect\citeauthoryear{Deng \& Zhang}{Deng \&
  Zhang}{2014}]{Deng_2014}
Deng W.,  Zhang B.,  2014, \mn@doi [\apj] {10.1088/0004-637x/785/2/112}, 785,
  112

\bibitem[\protect\citeauthoryear{{Deng} et~al.,}{{Deng}
  et~al.}{2022}]{2022arXiv220508737D}
{Deng} L.-T.,  et~al., 2022, arXiv e-prints, \href
  {https://ui.adsabs.harvard.edu/abs/2022arXiv220508737D} {p. arXiv:2205.08737}

\bibitem[\protect\citeauthoryear{{Drenkhahn}}{{Drenkhahn}}{2002}]{2002A&A...387..714D}
{Drenkhahn} G.,  2002, \mn@doi [\aap] {10.1051/0004-6361:20020390}, \href
  {https://ui.adsabs.harvard.edu/abs/2002A&A...387..714D} {387, 714}

\bibitem[\protect\citeauthoryear{{Drenkhahn} \& {Spruit}}{{Drenkhahn} \&
  {Spruit}}{2002}]{2002A&A...391.1141D}
{Drenkhahn} G.,  {Spruit} H.~C.,  2002, \mn@doi [\aap]
  {10.1051/0004-6361:20020839}, \href
  {https://ui.adsabs.harvard.edu/abs/2002A&A...391.1141D} {391, 1141}

\bibitem[\protect\citeauthoryear{{D’Elia}, {Siegel}  \& {Swift
  team}}{{D’Elia} et~al.}{2022}]{2022GCN.31966....1D}
{D’Elia} V.,  {Siegel} M.~H.,   {Swift team} 2022, GRB Coordinates Network,
  \href {https://ui.adsabs.harvard.edu/abs/2022GCN.31966....1D} {31966, 1}

\bibitem[\protect\citeauthoryear{{Gao} \& {Zhang}}{{Gao} \&
  {Zhang}}{2015}]{2015ApJ...801..103G}
{Gao} H.,  {Zhang} B.,  2015, \mn@doi [\apj] {10.1088/0004-637X/801/2/103},
  \href {https://ui.adsabs.harvard.edu/abs/2015ApJ...801..103G} {801, 103}

\bibitem[\protect\citeauthoryear{{Geng}, {Huang}, {Wu}, {Zhang}  \&
  {Zong}}{{Geng} et~al.}{2018}]{2018ApJS..234....3G}
{Geng} J.-J.,  {Huang} Y.-F.,  {Wu} X.-F.,  {Zhang} B.,   {Zong} H.-S.,  2018,
  \mn@doi [\apjs] {10.3847/1538-4365/aa9e84}, \href
  {https://ui.adsabs.harvard.edu/abs/2018ApJS..234....3G} {234, 3}

\bibitem[\protect\citeauthoryear{{Giannios}}{{Giannios}}{2006}]{2006A&A...457..763G}
{Giannios} D.,  2006, \mn@doi [\aap] {10.1051/0004-6361:20065000}, \href
  {https://ui.adsabs.harvard.edu/abs/2006A&A...457..763G} {457, 763}

\bibitem[\protect\citeauthoryear{{Giannios}}{{Giannios}}{2008}]{2008A&A...480..305G}
{Giannios} D.,  2008, \mn@doi [\aap] {10.1051/0004-6361:20079085}, \href
  {https://ui.adsabs.harvard.edu/abs/2008A&A...480..305G} {480, 305}

\bibitem[\protect\citeauthoryear{{Giannios} \& {Spruit}}{{Giannios} \&
  {Spruit}}{2005}]{2004Spectra}
{Giannios} D.,  {Spruit} H.~C.,  2005, \mn@doi [\aap]
  {10.1051/0004-6361:20047033}, \href
  {https://ui.adsabs.harvard.edu/abs/2005A&A...430....1G} {430, 1}

\bibitem[\protect\citeauthoryear{{Goodman}}{{Goodman}}{1986}]{1986Are}
{Goodman} J.,  1986, \mn@doi [\apjl] {10.1086/184741}, \href
  {https://ui.adsabs.harvard.edu/abs/1986ApJ...308L..47G} {308, L47}

\bibitem[\protect\citeauthoryear{{Gu}, {Liu}  \& {Lu}}{{Gu}
  et~al.}{2006}]{2006ApJ...643L..87G}
{Gu} W.-M.,  {Liu} T.,   {Lu} J.-F.,  2006, \mn@doi [\apjl] {10.1086/505140},
  \href {https://ui.adsabs.harvard.edu/abs/2006ApJ...643L..87G} {643, L87}

\bibitem[\protect\citeauthoryear{Hou et~al.,}{Hou et~al.}{2018}]{Hou_2018}
Hou S.-J.,  et~al., 2018, \mn@doi [\apj] {10.3847/1538-4357/aadc07}, 866, 13

\bibitem[\protect\citeauthoryear{{Ito} et~al.,}{{Ito}
  et~al.}{2013}]{2013ApJ...777...62I}
{Ito} H.,  et~al., 2013, \mn@doi [\apj] {10.1088/0004-637X/777/1/62}, \href
  {https://ui.adsabs.harvard.edu/abs/2013ApJ...777...62I} {777, 62}

\bibitem[\protect\citeauthoryear{{Janiuk}, {Yuan}, {Perna}  \& {Di
  Matteo}}{{Janiuk} et~al.}{2007}]{2007ApJ...664.1011J}
{Janiuk} A.,  {Yuan} Y.,  {Perna} R.,   {Di Matteo} T.,  2007, \mn@doi [\apj]
  {10.1086/518761}, \href
  {https://ui.adsabs.harvard.edu/abs/2007ApJ...664.1011J} {664, 1011}

\bibitem[\protect\citeauthoryear{{Kohri} \& {Mineshige}}{{Kohri} \&
  {Mineshige}}{2002}]{2002ApJ...577..311K}
{Kohri} K.,  {Mineshige} S.,  2002, \mn@doi [\apj] {10.1086/342166}, \href
  {https://ui.adsabs.harvard.edu/abs/2002ApJ...577..311K} {577, 311}

\bibitem[\protect\citeauthoryear{{Lazzati}, {Deich}, {Morsony}  \&
  {Workman}}{{Lazzati} et~al.}{2017}]{2017MNRAS.471.1652L}
{Lazzati} D.,  {Deich} A.,  {Morsony} B.~J.,   {Workman} J.~C.,  2017, \mn@doi
  [\mnras] {10.1093/mnras/stx1683}, \href
  {https://ui.adsabs.harvard.edu/abs/2017MNRAS.471.1652L} {471, 1652}

\bibitem[\protect\citeauthoryear{{Lei}, {Wang}, {Zhang}, {Gan}, {Zou}  \&
  {Xie}}{{Lei} et~al.}{2009}]{2009ApJ...700.1970L}
{Lei} W.~H.,  {Wang} D.~X.,  {Zhang} L.,  {Gan} Z.~M.,  {Zou} Y.~C.,   {Xie}
  Y.,  2009, \mn@doi [\apj] {10.1088/0004-637X/700/2/1970}, \href
  {https://ui.adsabs.harvard.edu/abs/2009ApJ...700.1970L} {700, 1970}

\bibitem[\protect\citeauthoryear{{Lei}, {Zhang}  \& {Liang}}{{Lei}
  et~al.}{2013}]{2013ApJ...765..125L}
{Lei} W.-H.,  {Zhang} B.,   {Liang} E.-W.,  2013, \mn@doi [\apj]
  {10.1088/0004-637X/765/2/125}, \href
  {https://ui.adsabs.harvard.edu/abs/2013ApJ...765..125L} {765, 125}

\bibitem[\protect\citeauthoryear{{Levinson} \& {Eichler}}{{Levinson} \&
  {Eichler}}{2003}]{2003ApJ...594L..19L}
{Levinson} A.,  {Eichler} D.,  2003, \mn@doi [\apjl] {10.1086/378487}, \href
  {https://ui.adsabs.harvard.edu/abs/2003ApJ...594L..19L} {594, L19}

\bibitem[\protect\citeauthoryear{{Liu}, {Liang}, {Gu}, {Zhao}, {Dai}  \&
  {Lu}}{{Liu} et~al.}{2010}]{2010A&A...516A..16L}
{Liu} T.,  {Liang} E.~W.,  {Gu} W.~M.,  {Zhao} X.~H.,  {Dai} Z.~G.,   {Lu}
  J.~F.,  2010, \mn@doi [\aap] {10.1051/0004-6361/200913447}, \href
  {https://ui.adsabs.harvard.edu/abs/2010A&A...516A..16L} {516, A16}

\bibitem[\protect\citeauthoryear{Lundman, Pe'er  \& Ryde}{Lundman
  et~al.}{2012}]{2013A}
Lundman C.,  Pe'er A.,   Ryde F.,  2012, \mn@doi [\mnras]
  {10.1093/mnras/sts219}, 428, 2430

\bibitem[\protect\citeauthoryear{Lü, Zou, Lei, Zhang, Wu, Wang, Liang  \&
  Lü}{Lü et~al.}{2012}]{L_2012}
Lü J.,  Zou Y.-C.,  Lei W.-H.,  Zhang B.,  Wu Q.,  Wang D.-X.,  Liang E.-W.,
  Lü H.-J.,  2012, \mn@doi [\apj] {10.1088/0004-637x/751/1/49}, 751, 49

\bibitem[\protect\citeauthoryear{{Malacaria}, {Meegan}  \& {Fermi GBM
  Team}}{{Malacaria} et~al.}{2022}]{2022GCN.31955....1M}
{Malacaria} C.,  {Meegan} C.,   {Fermi GBM Team} 2022, GRB Coordinates Network,
  \href {https://ui.adsabs.harvard.edu/abs/2022GCN.31955....1M} {31955, 1}

\bibitem[\protect\citeauthoryear{Matteo, Perna  \& Narayan}{Matteo
  et~al.}{2002}]{Di_Matteo_2002}
Matteo T.~D.,  Perna R.,   Narayan R.,  2002, \mn@doi [\apj] {10.1086/342832},
  579, 706

\bibitem[\protect\citeauthoryear{{Meng}, {Geng}  \& {Wu}}{{Meng}
  et~al.}{2022}]{2022MNRAS.509.6047M}
{Meng} Y.-Z.,  {Geng} J.-J.,   {Wu} X.-F.,  2022, \mn@doi [\mnras]
  {10.1093/mnras/stab3132}, \href
  {https://ui.adsabs.harvard.edu/abs/2022MNRAS.509.6047M} {509, 6047}

\bibitem[\protect\citeauthoryear{{M{\'e}sz{\'a}ros} \&
  {Rees}}{{M{\'e}sz{\'a}ros} \& {Rees}}{2000}]{2000ApJ...530..292M}
{M{\'e}sz{\'a}ros} P.,  {Rees} M.~J.,  2000, \mn@doi [\apj] {10.1086/308371},
  \href {https://ui.adsabs.harvard.edu/abs/2000ApJ...530..292M} {530, 292}

\bibitem[\protect\citeauthoryear{{Metzger}, {Piro}  \& {Quataert}}{{Metzger}
  et~al.}{2008}]{2008MNRAS.390..781M}
{Metzger} B.~D.,  {Piro} A.~L.,   {Quataert} E.,  2008, \mn@doi [\mnras]
  {10.1111/j.1365-2966.2008.13789.x}, \href
  {https://ui.adsabs.harvard.edu/abs/2008MNRAS.390..781M} {390, 781}

\bibitem[\protect\citeauthoryear{{Narayan}, {Piran}  \& {Kumar}}{{Narayan}
  et~al.}{2001}]{2001ApJ...557..949N}
{Narayan} R.,  {Piran} T.,   {Kumar} P.,  2001, \mn@doi [\apj]
  {10.1086/322267}, \href
  {https://ui.adsabs.harvard.edu/abs/2001ApJ...557..949N} {557, 949}

\bibitem[\protect\citeauthoryear{{Paczynski}}{{Paczynski}}{1986}]{1986ApJ308L43P}
{Paczynski} B.,  1986, \mn@doi [\apjl] {10.1086/184740}, \href
  {https://ui.adsabs.harvard.edu/abs/1986ApJ...308L..43P} {308, L43}

\bibitem[\protect\citeauthoryear{{Pe'er}}{{Pe'er}}{2008}]{2008ApJ...682..463P}
{Pe'er} A.,  2008, \mn@doi [\apj] {10.1086/588136}, \href
  {https://ui.adsabs.harvard.edu/abs/2008ApJ...682..463P} {682, 463}

\bibitem[\protect\citeauthoryear{{Pe'er}, {M{\'e}sz{\'a}ros}  \&
  {Rees}}{{Pe'er} et~al.}{2005}]{2005ApJ...635..476P}
{Pe'er} A.,  {M{\'e}sz{\'a}ros} P.,   {Rees} M.~J.,  2005, \mn@doi [\apj]
  {10.1086/497360}, \href
  {https://ui.adsabs.harvard.edu/abs/2005ApJ...635..476P} {635, 476}

\bibitem[\protect\citeauthoryear{Pe'er, Meszaros  \& Rees}{Pe'er
  et~al.}{2006}]{Pe_er_2006}
Pe'er A.,  Meszaros P.,   Rees M.~J.,  2006, \mn@doi [\apj] {10.1086/507595},
  652, 482

\bibitem[\protect\citeauthoryear{{Pe'er}, {Ryde}, {Wijers}, {M{\'e}sz{\'a}ros}
  \& {Rees}}{{Pe'er} et~al.}{2007}]{2007ApJ...664L...1P}
{Pe'er} A.,  {Ryde} F.,  {Wijers} R. A.~M.~J.,  {M{\'e}sz{\'a}ros} P.,   {Rees}
  M.~J.,  2007, \mn@doi [\apjl] {10.1086/520534}, \href
  {https://ui.adsabs.harvard.edu/abs/2007ApJ...664L...1P} {664, L1}

\bibitem[\protect\citeauthoryear{{Pe'er}, {Barlow}, {O'Mahony}, {Margutti},
  {Ryde}, {Larsson}, {Lazzati}  \& {Livio}}{{Pe'er}
  et~al.}{2015}]{2015ApJ...813..127P}
{Pe'er} A.,  {Barlow} H.,  {O'Mahony} S.,  {Margutti} R.,  {Ryde} F.,
  {Larsson} J.,  {Lazzati} D.,   {Livio} M.,  2015, \mn@doi [\apj]
  {10.1088/0004-637X/813/2/127}, \href
  {https://ui.adsabs.harvard.edu/abs/2015ApJ...813..127P} {813, 127}

\bibitem[\protect\citeauthoryear{{Popham}, {Woosley}  \& {Fryer}}{{Popham}
  et~al.}{1999}]{1999ApJ...518..356P}
{Popham} R.,  {Woosley} S.~E.,   {Fryer} C.,  1999, \mn@doi [\apj]
  {10.1086/307259}, \href
  {https://ui.adsabs.harvard.edu/abs/1999ApJ...518..356P} {518, 356}

\bibitem[\protect\citeauthoryear{{Preece}, {Briggs}, {Mallozzi}, {Pendleton},
  {Paciesas}  \& {Band}}{{Preece} et~al.}{1998}]{Preece_1998}
{Preece} R.~D.,  {Briggs} M.~S.,  {Mallozzi} R.~S.,  {Pendleton} G.~N.,
  {Paciesas} W.~S.,   {Band} D.~L.,  1998, \mn@doi [\apjl] {10.1086/311644},
  \href {https://ui.adsabs.harvard.edu/abs/1998ApJ...506L..23P} {506, L23}

\bibitem[\protect\citeauthoryear{{Rees} \& {Meszaros}}{{Rees} \&
  {Meszaros}}{1994}]{1994ApJ...430L..93R}
{Rees} M.~J.,  {Meszaros} P.,  1994, \mn@doi [\apjl] {10.1086/187446}, \href
  {https://ui.adsabs.harvard.edu/abs/1994ApJ...430L..93R} {430, L93}

\bibitem[\protect\citeauthoryear{{Rees} \& {M{\'e}sz{\'a}ros}}{{Rees} \&
  {M{\'e}sz{\'a}ros}}{2005}]{2005Dissipative}
{Rees} M.~J.,  {M{\'e}sz{\'a}ros} P.,  2005, \mn@doi [\apj] {10.1086/430818},
  \href {https://ui.adsabs.harvard.edu/abs/2005ApJ...628..847R} {628, 847}

\bibitem[\protect\citeauthoryear{{Rossi}, {Beloborodov}  \& {Rees}}{{Rossi}
  et~al.}{2006}]{2006MNRAS.369.1797R}
{Rossi} E.~M.,  {Beloborodov} A.~M.,   {Rees} M.~J.,  2006, \mn@doi [\mnras]
  {10.1111/j.1365-2966.2006.10417.x}, \href
  {https://ui.adsabs.harvard.edu/abs/2006MNRAS.369.1797R} {369, 1797}

\bibitem[\protect\citeauthoryear{{Ryde} et~al.,}{{Ryde}
  et~al.}{2010}]{2010ApJ...709L.172R}
{Ryde} F.,  et~al., 2010, \mn@doi [\apjl] {10.1088/2041-8205/709/2/L172}, \href
  {https://ui.adsabs.harvard.edu/abs/2010ApJ...709L.172R} {709, L172}

\bibitem[\protect\citeauthoryear{{Ryde} et~al.,}{{Ryde}
  et~al.}{2011}]{2011Observational}
{Ryde} F.,  et~al., 2011, \mn@doi [\mnras] {10.1111/j.1365-2966.2011.18985.x},
  \href {https://ui.adsabs.harvard.edu/abs/2011MNRAS.415.3693R} {415, 3693}

\bibitem[\protect\citeauthoryear{{Scargle}, {Norris}, {Jackson}  \&
  {Chiang}}{{Scargle} et~al.}{2013}]{2013ApJ...764..167S}
{Scargle} J.~D.,  {Norris} J.~P.,  {Jackson} B.,   {Chiang} J.,  2013, \mn@doi
  [\apj] {10.1088/0004-637X/764/2/167}, \href
  {https://ui.adsabs.harvard.edu/abs/2013ApJ...764..167S} {764, 167}

\bibitem[\protect\citeauthoryear{{Song} \& {Meng}}{{Song} \&
  {Meng}}{2022}]{2022MNRAS.512.5693S}
{Song} X.-Y.,  {Meng} Y.-Z.,  2022, \mn@doi [\mnras] {10.1093/mnras/stac839},
  \href {https://ui.adsabs.harvard.edu/abs/2022MNRAS.512.5693S} {512, 5693}

\bibitem[\protect\citeauthoryear{{Song}, {Zhang}, {Zhang}, {Xiong}  \&
  {Song}}{{Song} et~al.}{2022}]{2022arXiv220409430S}
{Song} X.-Y.,  {Zhang} S.-N.,  {Zhang} S.,  {Xiong} S.-L.,   {Song} L.-M.,
  2022, \mn@doi [\apj] {10.3847/1538-4357/ac6b33}, \href
  {https://ui.adsabs.harvard.edu/abs/2022ApJ...931..112S} {931, 112}

\bibitem[\protect\citeauthoryear{{Thompson}}{{Thompson}}{1994}]{1994A}
{Thompson} C.,  1994, \mn@doi [\mnras] {10.1093/mnras/270.3.480}, \href
  {https://ui.adsabs.harvard.edu/abs/1994MNRAS.270..480T} {270, 480}

\bibitem[\protect\citeauthoryear{{Uhm} \& {Zhang}}{{Uhm} \&
  {Zhang}}{2014}]{2014NatPh..10..351U}
{Uhm} Z.~L.,  {Zhang} B.,  2014, \mn@doi [Nature Physics] {10.1038/nphys2932},
  \href {https://ui.adsabs.harvard.edu/abs/2014NatPh..10..351U} {10, 351}

\bibitem[\protect\citeauthoryear{{Ukwatta} et~al.,}{{Ukwatta}
  et~al.}{2010}]{2010ApJ...711.1073U}
{Ukwatta} T.~N.,  et~al., 2010, \mn@doi [\apj] {10.1088/0004-637X/711/2/1073},
  \href {https://ui.adsabs.harvard.edu/abs/2010ApJ...711.1073U} {711, 1073}

\bibitem[\protect\citeauthoryear{{Vurm}, {Beloborodov}  \& {Poutanen}}{{Vurm}
  et~al.}{2011}]{2010Radiative}
{Vurm} I.,  {Beloborodov} A.~M.,   {Poutanen} J.,  2011, \mn@doi [\apj]
  {10.1088/0004-637X/738/1/77}, \href
  {https://ui.adsabs.harvard.edu/abs/2011ApJ...738...77V} {738, 77}

\bibitem[\protect\citeauthoryear{{Wang} et~al.,}{{Wang}
  et~al.}{2021}]{2021ApJ...922..237W}
{Wang} X.~I.,  et~al., 2021, \mn@doi [\apj] {10.3847/1538-4357/ac29bd}, \href
  {https://ui.adsabs.harvard.edu/abs/2021ApJ...922..237W} {922, 237}

\bibitem[\protect\citeauthoryear{{Wang}, {Zheng}  \& {Jin}}{{Wang}
  et~al.}{2022}]{2022arXiv220508427W}
{Wang} Y.,  {Zheng} T.-C.,   {Jin} Z.-P.,  2022, arXiv e-prints, \href
  {https://ui.adsabs.harvard.edu/abs/2022arXiv220508427W} {p. arXiv:2205.08427}

\bibitem[\protect\citeauthoryear{{Wei}, {Wu}  \& {Melia}}{{Wei}
  et~al.}{2016}]{2016MNRAS.463.1144W}
{Wei} J.-J.,  {Wu} X.-F.,   {Melia} F.,  2016, \mn@doi [\mnras]
  {10.1093/mnras/stw2057}, \href
  {https://ui.adsabs.harvard.edu/abs/2016MNRAS.463.1144W} {463, 1144}

\bibitem[\protect\citeauthoryear{{Yi}, {Lei}, {Zhang}, {Dai}, {Wu}  \&
  {Liang}}{{Yi} et~al.}{2017}]{YI20171}
{Yi} S.-X.,  {Lei} W.-H.,  {Zhang} B.,  {Dai} Z.-G.,  {Wu} X.-F.,   {Liang}
  E.-W.,  2017, \mn@doi [Journal of High Energy Astrophysics]
  {10.1016/j.jheap.2017.01.001}, \href
  {https://ui.adsabs.harvard.edu/abs/2017JHEAp..13....1Y} {13, 1}

\bibitem[\protect\citeauthoryear{{Zalamea} \& {Beloborodov}}{{Zalamea} \&
  {Beloborodov}}{2011}]{2011MNRAS.410.2302Z}
{Zalamea} I.,  {Beloborodov} A.~M.,  2011, \mn@doi [\mnras]
  {10.1111/j.1365-2966.2010.17600.x}, \href
  {https://ui.adsabs.harvard.edu/abs/2011MNRAS.410.2302Z} {410, 2302}

\makeatother
\end{thebibliography}




\appendix
\section{APPENDIX}
\subsection{Modeling}\label{sec:4funcs}
\subsubsection{BAND function}
The BAND function has four free parameters: low and high energy spectral indices, denoted as $\alpha$ and $\beta$ respectively, the peak energy of $\nu F_{\nu}$ spectrum, denoted as $E_{\rm p}$, and amplitude, as shown in Equation~\ref{func:BAND} ($N(E)$ is in units of ph cm$^{-2}$ s$^{-1}$ keV$^{-1}$, the same below).
\begin{equation}\label{func:BAND}
\begin{split}
&  N_{\rm BAND }(E) = \\
& A \begin{cases}
\biggl(\frac{E}{100 \ \rm keV }\biggr)^{\alpha} \exp \biggl[- \frac{ (\alpha +2) E}{ E_{\rm p} } \biggr], \ E \geq \frac{ (\alpha - \beta) \
E_{\rm p} } { \alpha +2} \\
\biggl( \frac{E}{ 100 \ \rm keV } \biggr)^{ \beta } \exp (\beta -\alpha) \biggl[ \frac{(\alpha-\beta ) E_{\rm p}}{100 \ \rm keV \ (\alpha
+2)} \biggr]^{\alpha-\beta }, \\ E < \frac{(\alpha -\beta ) \ E_{\rm p}}{\alpha +2}.
\end{cases}
\end{split}
\end{equation}

\subsubsection{CPL model}
CPL model is a subset of BAND model if $\beta$ is very small and the part of $E < \frac{(\alpha -\beta ) \ E_{\rm p}}{\alpha +2}$ of BAND model is ignored. There are three parameters in CPL model: the amplitude A, the lower energy index $\alpha$, and the $\nu F_{\nu}$ peak energy, $E_{\rm p}$, as shown in Equation~\ref{func:CPL}.
\begin{equation}\label{func:CPL}
N_{\rm COMP}(E) = A \ \Biggl(\frac{E}{1 keV}\Biggr)^{\alpha} \exp \Biggl[ -\frac{(\alpha+2) \ E}{E_{\rm p}}  \Biggr].
\end{equation}
\subsubsection{mBB model}
the photon spectrum of the mBB can be formulated as
\begin{equation}
N(E)=\frac{8.0525(m+1)K}{\left[ \left( \frac{T_{\max }}{T_{\min }}\right)
^{m+1}-1\right] }\left( \frac{kT_{\min }}{\mathrm{keV}}\right) ^{-2}I(E),
\end{equation}%
where
\begin{equation}
I(E)=\left( \frac{E}{kT_{\min }}\right) ^{m-1}\int_{\frac{E}{kT_{\max }}}^{%
\frac{E}{kT_{\min }}}\frac{x^{2-m}}{e^{x}-1}dx,
\end{equation}%
and $x=E/kT$;  $K=L_{39}/D^2_{L,\mathrm{10 kpc}}$
is defined by the blackbody luminosity $L$ in units of $10^{39}$ erg s$^{-1}$
in the GRB host galaxy frame and the luminosity distance $D_{L}$ in units of
10 kpc; $m$ is the power-law index of the distribution, and the temperature
ranges from the minimum $T_{\min }$ to the maximum $T_{\max }$.
\subsubsection{NDP of pure fireball}
The time-averaged flux $F_{\rm v}(E_{\rm obs}, r_0, \eta_0, p, \theta_{\rm c}, \theta_{\rm v}, L, z)$ of the observed energy $E_{\rm obs}$ is defined as  
\begin{equation}
\begin{split}
 &F_{\rm v}(E_{\rm obs}, r_0, \eta_0, p, \theta_{\rm c}, \theta_{\rm v}, L, z), 
\end{split}    
\end{equation}
where $r_0$ is the acceleration radius measured at the base of the outflow, which is the radius where the
acceleration of plasma to relativistic (kinetic) motion begins~\citep{2013A,2015ApJ...813..127P}. In the angle-dependent baryon loading parameter profile $\eta(\theta)\sim\frac{\eta_{0}}{
((\theta /\theta_{\rm c})^{2p}+1)^{1/2}}$, $\theta$ is the angle measured from the jet axis, $p$ is the power-law index and $\eta_0=\eta (\theta=0)$. If the emission is from the saturated regime ($R_{\rm ph} \geq R_{\rm s}$), $\Gamma=\eta$, otherwise, $\Gamma=R_{\rm ph}/r_0$. $\theta_{\rm v}$ is the line of sight (LOS) measured from the jet axis. $\theta_{\rm c}$ is the half-opening angle for the jet core. $L$ is the total outflow luminosity.
 $z$ is the redshift. The details are described in \cite{2022arXiv220409430S} and the references therein.

$R_{\rm ph}$ in different regimes is defined as
\begin{equation}\label{eq:Rph}
R_{\text{ph}}=\\
\begin{split}
&\begin{cases}
\left(\frac{\sigma _{\text{T}}}{6m_{\text{p}}c}%
\frac{d\dot{M}}{d\Omega }r_{0}^{2}\right) ^{1/3}\text{, } R_{\rm ph} \ll R_{\rm s} \text{, }\\
\left(\frac{\sigma _{\text{T}}}{2m_{\text{p}}c}%
\frac{d\dot{M}}{d\Omega }r_{0}^{2}\right) ^{1/3}\text{, } R_{\rm ph} \lesssim R_{\rm s} \text{, } \\
\frac{1}{(1+\beta )\beta \eta ^{2}}\frac{\sigma _{\text{T}}
}{m_{\text{p}}c}\frac{d\dot{M}}{d\Omega } \text{, }   R_{\rm ph} \geq R_{\rm s},\\
\end{cases}
\end{split}
\end{equation}
and $\Gamma$ in different regimes are determined to be
\begin{equation}\label{eq:gamma}
\Gamma=\begin{split}
&\begin{cases}
R_{\rm ph}/r_0\text{,  }R_{\rm ph} < R_{\rm s} \text{, }\\
\eta \text{,  } R_{\rm ph} \geq R_{\rm s},\\
\end{cases}
\end{split}
\end{equation}
where $\beta$ is the velocity, and $d\dot{M}(\theta )/d\Omega =L/4\pi c^{2}\eta (\theta )$ is the angle-dependent mass outflow rate per solid angle, $m_{\rm p}$ is the mass of the proton, $c$ is the light speed, and $\sigma_{\rm T}$ is electron Thomson cross section. In this analysis, we assume the jet is viewed on-axis to perform the fit, thus, $\theta_{\rm v}=0$ for this bright burst. $\theta_{\rm v}=0$ is a good approximation for small $\theta_{\rm v}$, i.e., $\theta_{\rm v}$ is much less than the jet opening angle, as discussed in \cite{2013A}. There is a critical angle $\theta_{\rm cri}$~\citep{2022MNRAS.512.5693S}, and the regime turns from unsaturated to saturated for $\theta>\theta_{\rm cri}$, where $\theta$ is the angle measured from the jet axis. $\theta_{\rm cri}$ is determined by the extracted parameters from the fit. If $\theta_{\rm cri} \gtrsim 5/\eta_0$, the prompt emission is dominantly from the unsaturated regime~\citep{2013A}.
In the combined model NDP +IC, the non-thermal component in the higher energy band is described as a CPL function with the amplitude $C$, the power law photon index $B$, and the $\nu F_{\nu}$ peak energy, $E_{\rm p}$, as shown in 
\begin{equation}\label{func:CPL2}
N_{\rm CPL}(E) = C (\frac{E}{1  \rm{keV}})^{B} \exp \Biggl[ -\frac{(B+2) \ E}{E_{\rm p}}  \Biggr].
\end{equation}
 A combination model of photospheric and non-thermal component is used in the fitting, labeled as NDP+CPL.
\subsubsection{Synchrotron Model}
In synchrotron model \citep[e.g.,][]{2014NatPh..10..351U, 2018ApJS..234....3G}, a group of electrons, which obey a power-law
distribution, i.e., $Q(\gamma _{\mathrm{e}}^{\prime },t^{\prime
})=Q_{0}(t^{\prime
})(\gamma _{\mathrm{e}}^{\prime }/\gamma _{\mathrm{m}}^{\prime })^{-p}$ for $%
\gamma _{\mathrm{e}}^{\prime }>\gamma _{\mathrm{m}}^{\prime }$, are injected
in the relativistically moving shell of Lorentz factor $\Gamma$. Here, $%
Q_{0}$ is related to the injection rate $N_{\mathrm{inj}}^{\prime }$ by $N_{%
\mathrm{inj}}^{\prime }=\int_{\gamma _{\mathrm{m}}^{\prime }}^{\gamma _{%
\mathrm{max}}^{\prime }}Q(\gamma _{\mathrm{e}}^{\prime },t^{\prime })d\gamma
_{\mathrm{e}}^{\prime }$, where $\gamma _{\mathrm{max}}^{\prime }$ is the
maximum Lorentz factor of electrons, and $\gamma _{\mathrm{m}}^{\prime}$ is the minimum Lorentz factor of electrons. For an electron of $\gamma _{\mathrm{e}%
}^{\prime }$, it would lose energy by synchrotron radiation, of which the
cooling rate is
\begin{equation}
\dot{\gamma}_{\mathrm{e}}^{\prime }=-\frac{\sigma _{T}B^{\prime 2}\gamma _{%
\mathrm{e}}^{\prime 2}}{6\pi m_{\mathrm{e}}c},  \label{eq:syn}
\end{equation}%
where $B^{\prime }$ is the magnetic field in the co-moving frame.

Considering a conical jet, the co-moving magnetic field in the jet would
decay with radius as
\begin{equation}
B^{\prime }=B_{0}^{\prime }\left( \frac{R}{R_{0}}\right) ^{-b},
\end{equation}%
where $B_{0}^{\prime }$ is the magnetic strength at $R_{0}$, and $R_{0}$ is
the radius where the jet begins to emit the first photon observed by us. $R_{0}=2\Gamma ^{2}c\times 1~\mathrm{s}$, and denote
observer-frame time since the first electron injection as $\hat{t}$ (in
units of s) for an emission episode. Only the
emission from the region just near the LOS and treat this small region as a
uniform jet is considered. So relevant parameters in our calculation describe properties
of the region near the LOS, rather than those of the jet axis. This
treatment enables us to simplify the calculation and focus on properties of
the region near the LOS. The flux density of this model (in unit of mJy) is in the form 
\begin{equation}
 F_{\nu}(E) = F_{\nu}(E;\Gamma,p, \gamma _{\mathrm{m}}^{\prime },R_{inj}^{0},q,B_{0}^{\prime },b,\hat{t}).
\end{equation}
The fit can constrain eight parameters, including $\Gamma$, the power-law index of the electron spectrum $p$, the minimum Lorentz factor of electrons $\gamma _{\mathrm{m}}^{\prime}$, the normalized injection rate of electrons $R_{inj}^{0}$, the power-law index of the injection rate $q$, the initial magnetic filed $B_{0}^{\prime }$, the decaying factor of the magnetic field $b$ and the time at which electrons begin to radiate in the observer frame $\hat{t}$. Note that the spectrum does not depend on the redshift $z$. In the fitting procedure, z is fixed at 1.
\subsection{Time-resolved fit results with NDP+CL model}\label{sec:TRspectra}
\begin{figure*}
\begin{center}
 \centering
   \includegraphics[width=0.5\textwidth]{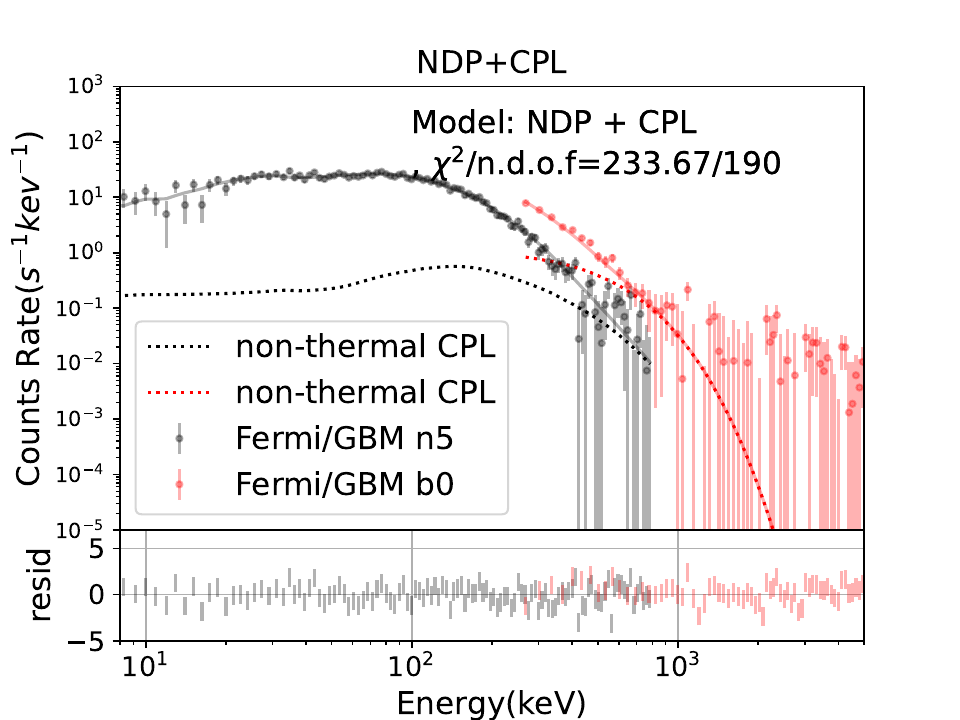}\put(-200,160){(a)[-0.05,2.22]s}
   \includegraphics[width=0.5\textwidth]{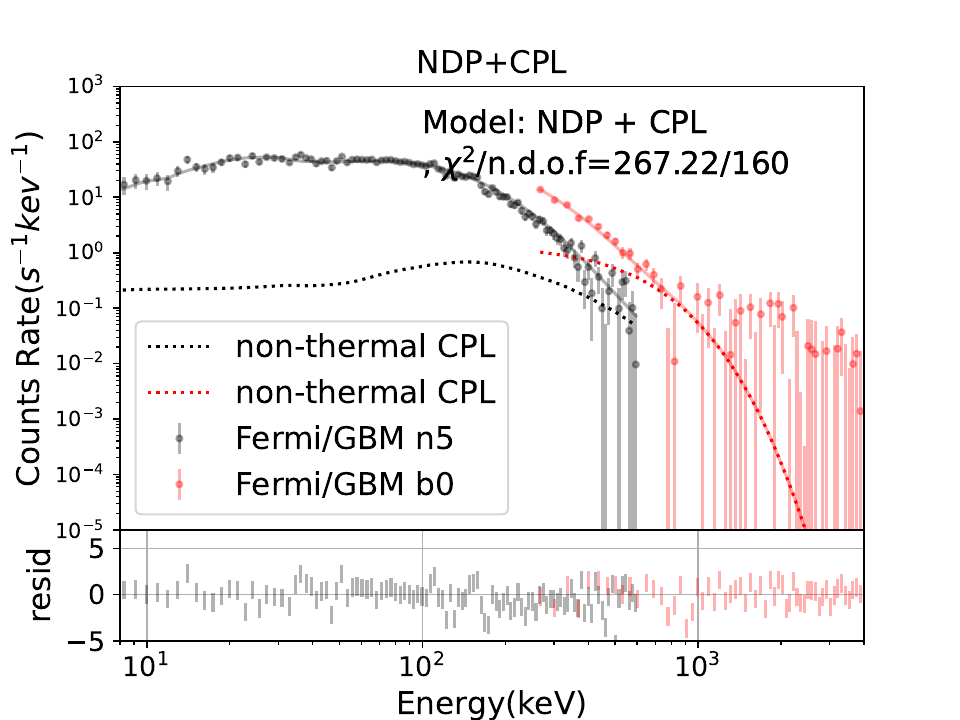}\put(-200,160){(b)[2.22,3.42]s}   \\
   \includegraphics[width=0.5\textwidth]{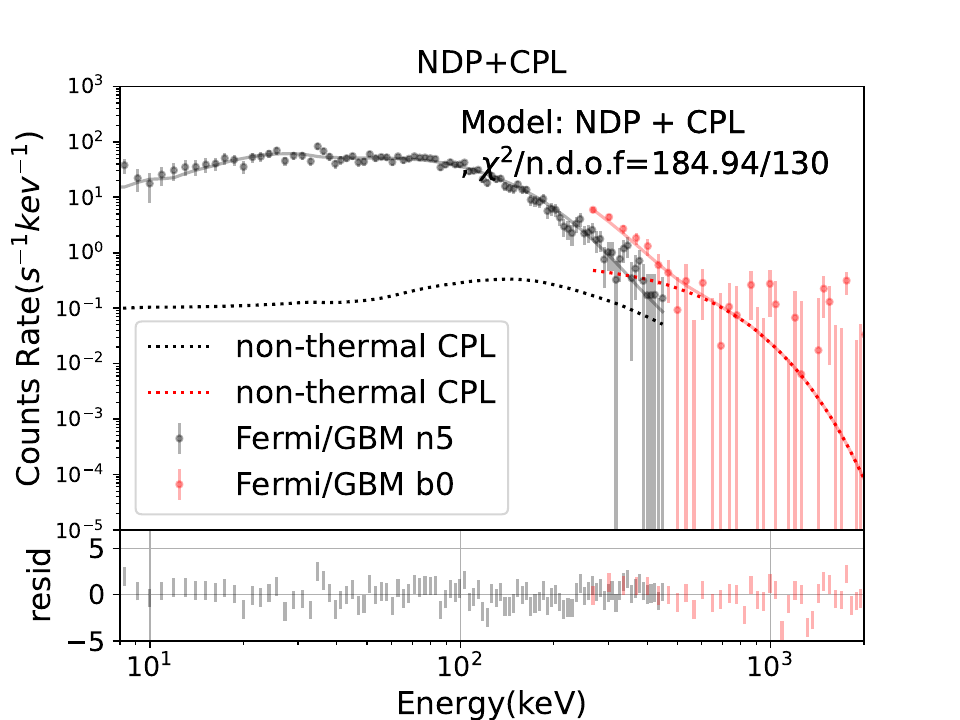}\put(-200,160){(c)[3.42,3.92]s}
   \includegraphics[width=0.5\textwidth]{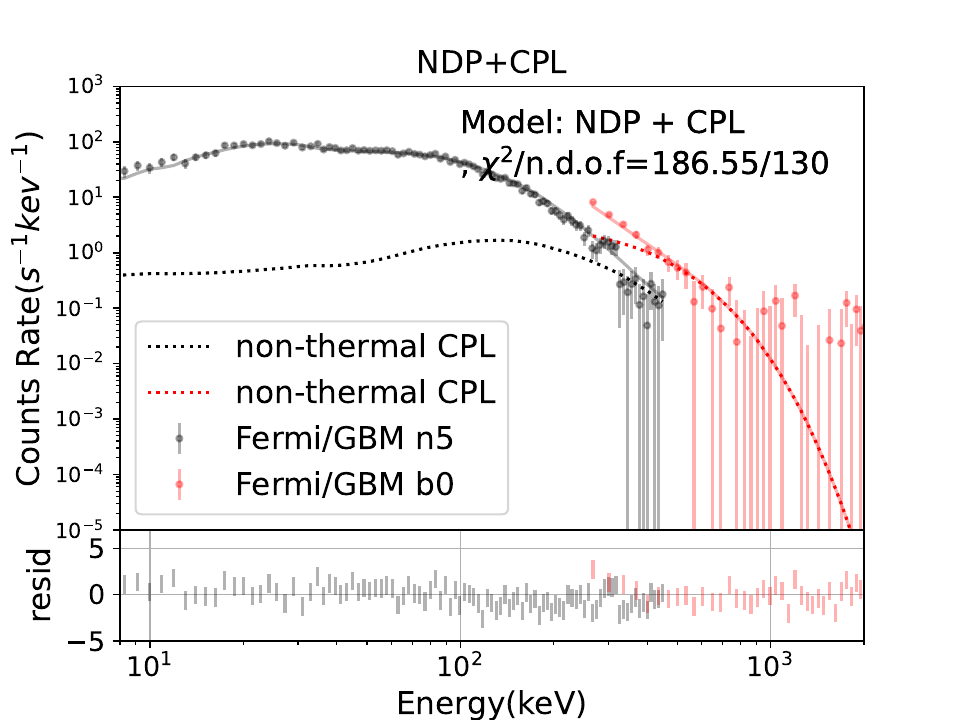}\put(-200,160){(d)[3.92,5.07]s} \\ 
   \includegraphics[width=0.5\textwidth]{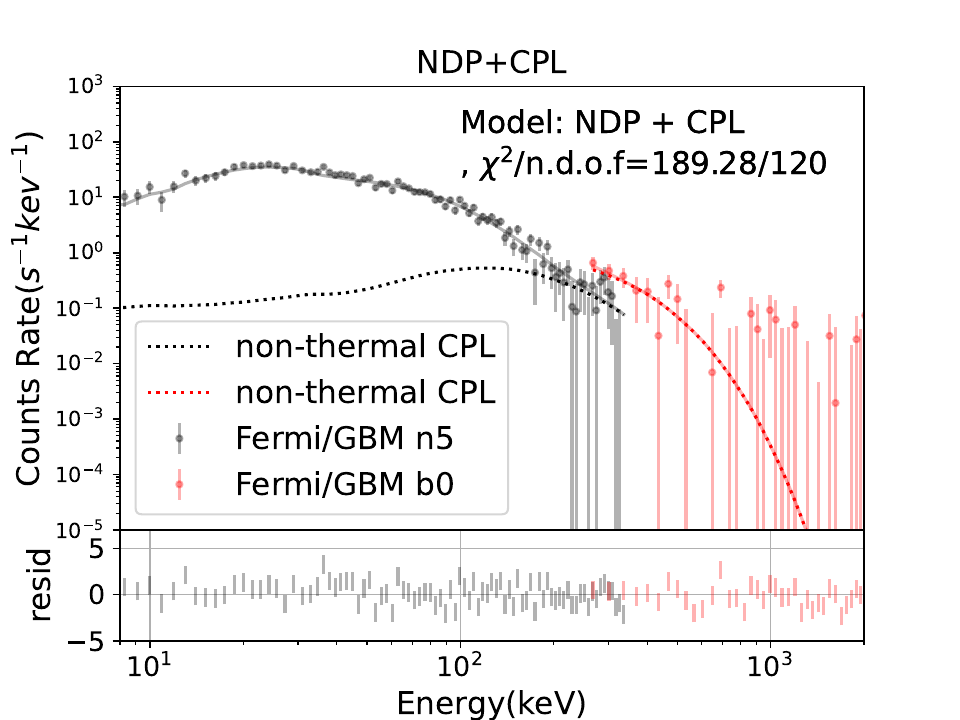}\put(-200,160){(d)[5.07,7.79]s} 
   \includegraphics[width=0.5\textwidth]{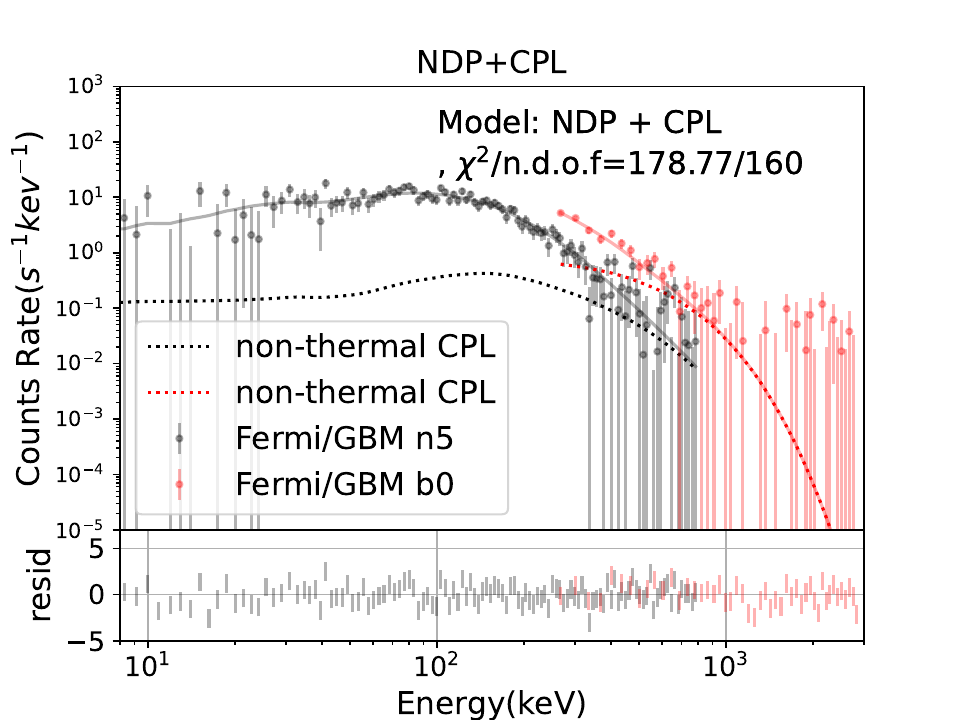}\put(-200,160){(e)[-0.05,0.95]s}  \\
\caption{(a)-(d) are the time-resolved spectra and the fit results of BBlocks binning with float parameters. (e) is for the first 1 s. \label{fig:TRres_floatpara} }
\end{center}
\end{figure*}
\clearpage
\subsection{Time-resolved fit results with mBB model}\label{sec:TRspectra2}
\begin{figure*}
\begin{center}
 \centering
   \includegraphics[width=0.5\textwidth]{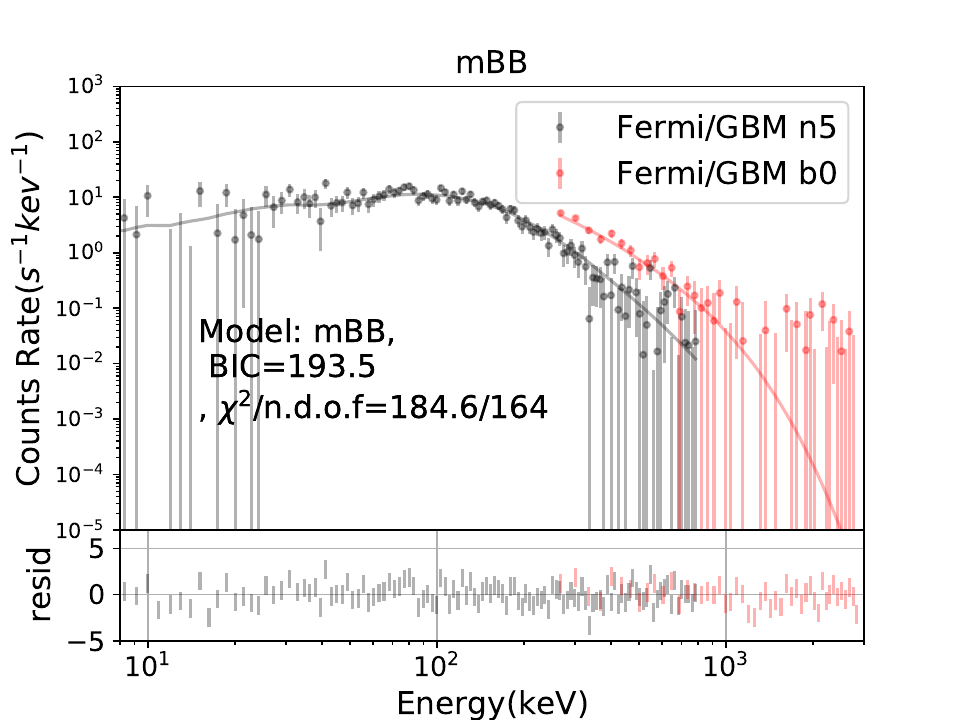}\put(-180,160){(a)[-0.05,0.95]s}
   \includegraphics[width=0.5\textwidth]{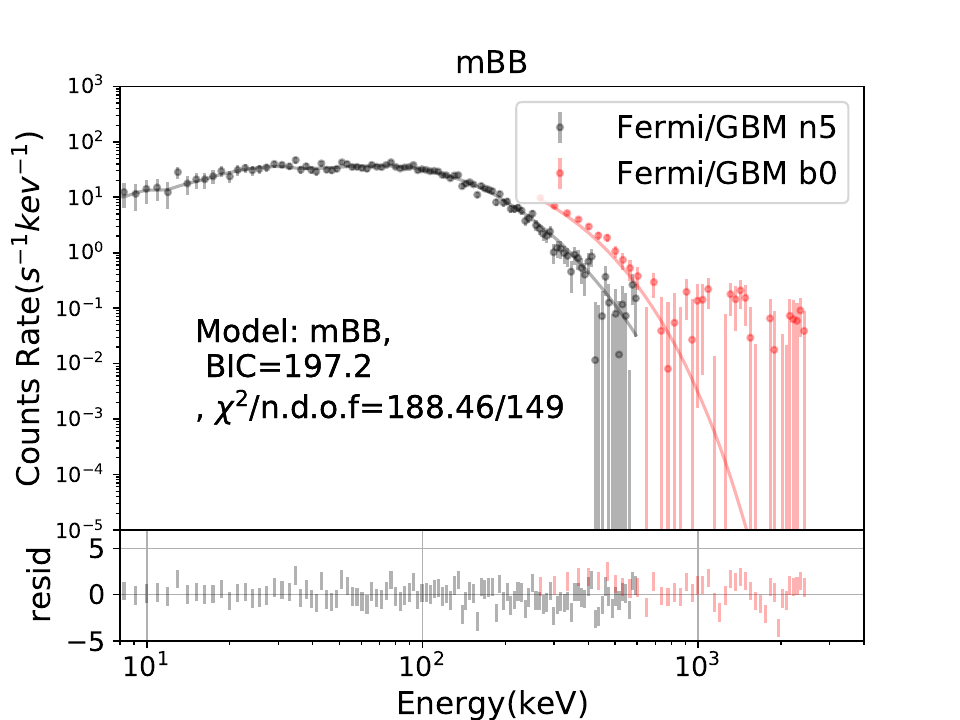}\put(-180,160){(b)[0.95,1.95]s}  \\ 
   \includegraphics[width=0.5\textwidth]{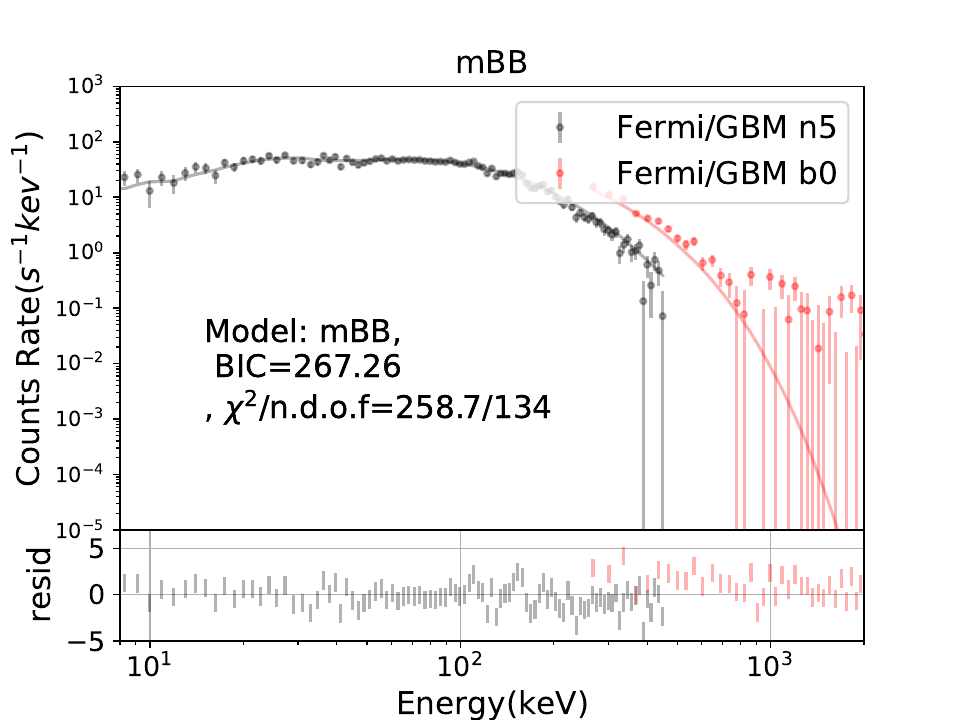}\put(-180,160){(c)[1.95,2.95]s}
   \includegraphics[width=0.5\textwidth]{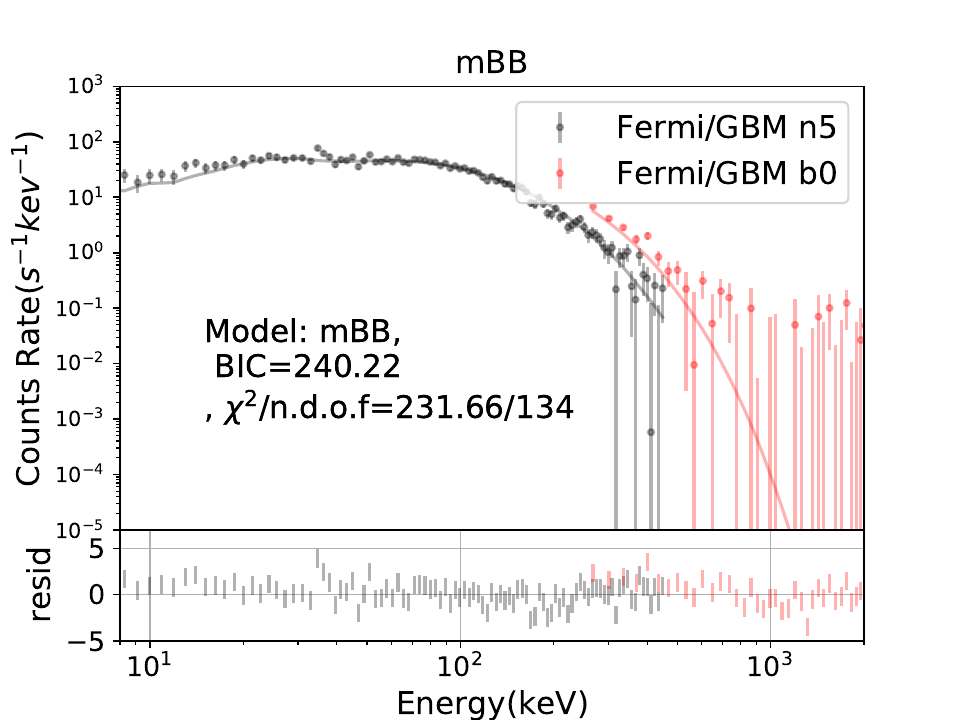}\put(-180,160){(d)[2.95,3.95]s} \\
   \includegraphics[width=0.5\textwidth]{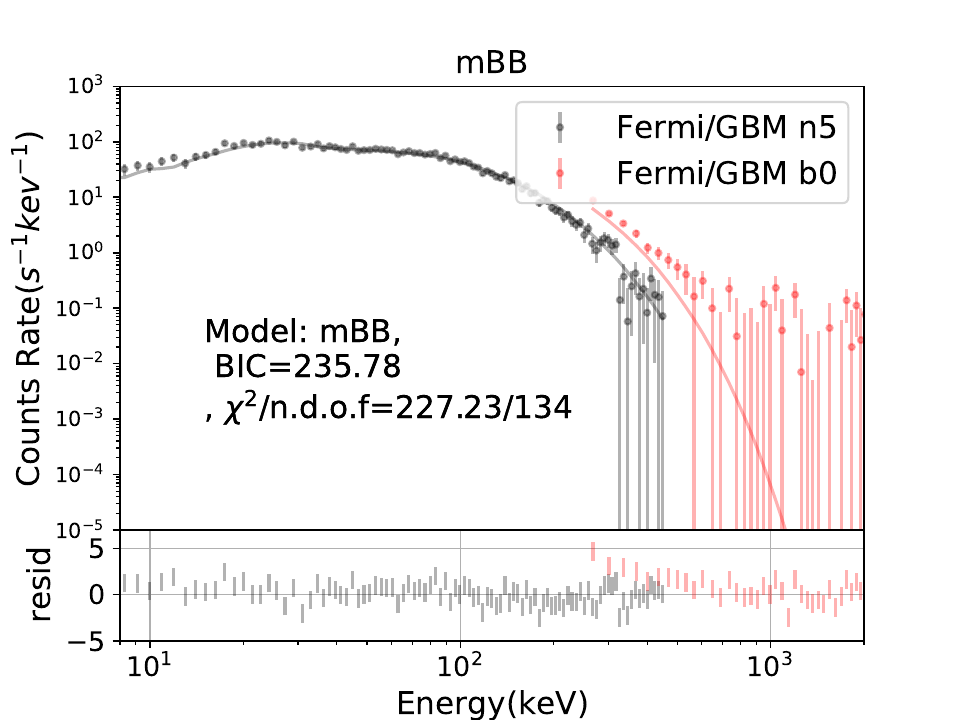}\put(-180,160){(d)[3.95,4.95]s} 
   \includegraphics[width=0.5\textwidth]{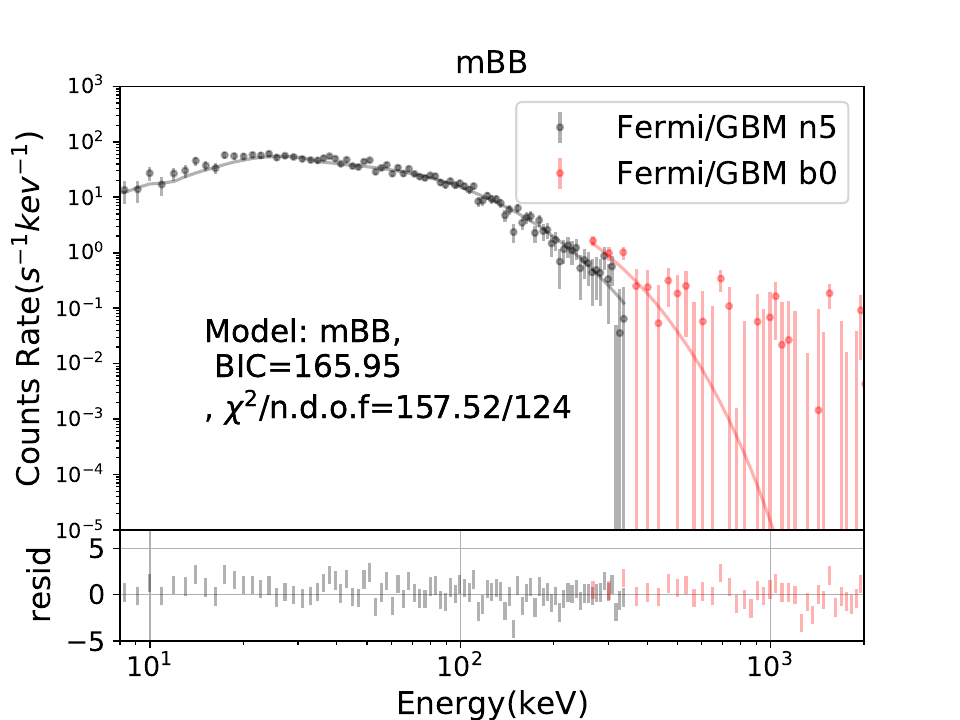}\put(-180,160){(e)[4.95,5.95]s}  
\caption{(a)-(f) are the time-resolved spectra and the fit results of CC binning with float parameters with mBB. \label{fig:TRres_floatpara2} }
\end{center}
\end{figure*}
\begin{figure*}
\begin{center}
 \centering   
   \includegraphics[width=0.5\textwidth]{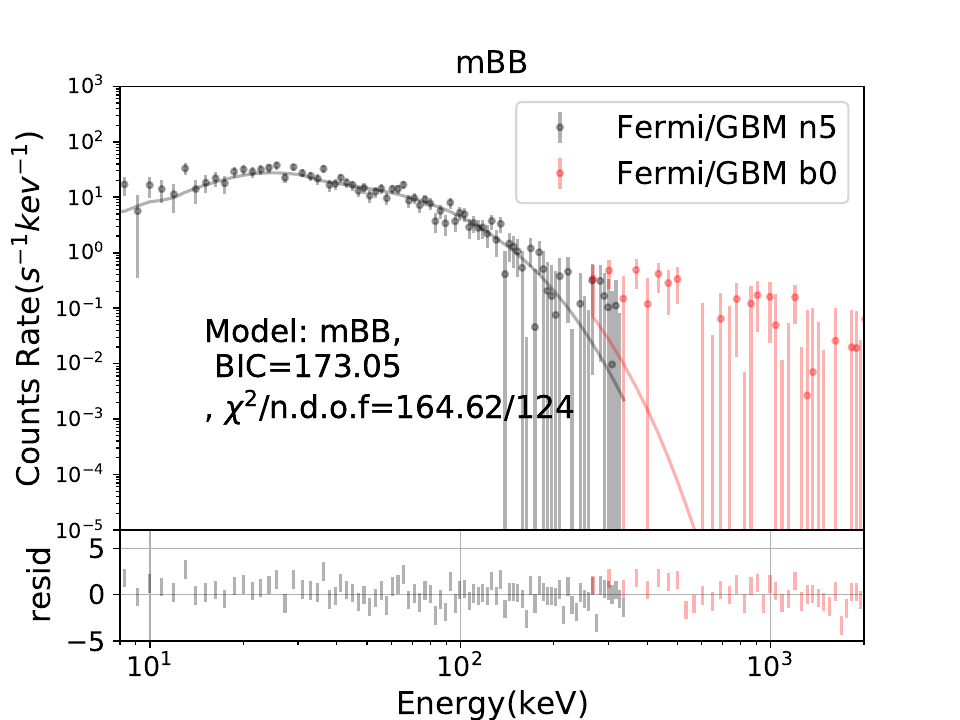}\put(-180,160){(f)[-5.95,6.95]s}  
\center{Figure~\ref{fig:TRres_floatpara2} (Continued.)}
\end{center}
\end{figure*}




\bsp	
\label{lastpage}
\end{document}